\documentclass[11pt]{article}

\usepackage{graphicx,latexsym,amssymb,amsmath}

\usepackage{cite} 

\def\truered{}
\def\green{}
\def\black{}
\newcommand{\Fig}[1]{Fig.\ref{#1}}

\newcommand{\edo}{\end{document}}

\newcommand{\fractextnp}[2]{#1/#2}

\newcommand{\pic}[2]{\includegraphics[scale=#1]{#2}}

\newtheorem{itlemma}{Lemma}[section] 

\newtheorem{itremark}[itlemma]{Remark}

\newenvironment{myremark}{\begin{itremark}\rm}{\end{itremark}} 

\newcommand{\be}[1]{\begin{equation}\label{#1}}
\newcommand{\ee}{\end{equation}}
\newcommand{\bl}[1]{\begin{lemma}\label{#1}}
\newcommand{\ble}[1]{\begin{lemmaex}\label{#1}}
\newcommand{\br}[1]{\begin{myremark}\label{#1}}
\newcommand{\bt}[1]{\begin{theorem}\label{#1}}
\newcommand{\bd}[1]{\begin{definition}\label{#1}}
\newcommand{\bp}[1]{\begin{proposition}\label{#1}}
\newcommand{\bc}[1]{\begin{corollary}\label{#1}}
\newcommand{\bfact}[1]{\begin{fact}\label{#1}}
\newcommand{\ber}[1]{\begin{exercise}\label{#1}}
\newcommand{\bex}[1]{\begin{example}\label{#1}}
\newcommand{\bem}[1]{\begin{example}\label{#1}}  
\newcommand{\ec}{\mybox\end{corollary}}
\newcommand{\efact}{\mybox\end{fact}}
\newcommand{\eer}{\mybox\end{exercise}}
\newcommand{\eex}{\mybox\end{example}}
\newcommand{\eem}{\mybox\end{example}}
\newcommand{\el}{\mybox\end{lemma}}
\newcommand{\ele}{\mybox\end{lemmaex}}
\newcommand{\er}{\mybox\end{myremark}}
\newcommand{\et}{\qed\end{theorem}}
\newcommand{\ed}{\mybox\end{definition}}
\newcommand{\ep}{\mybox\end{proposition}}
\newcommand{\epr}{\end{proof}}
\newcommand{\bpr}{\begin{proof}}

\newcommand{\ecs}{\end{corollary}}
\newcommand{\eers}{\end{exercise}}
\newcommand{\eexs}{\end{example}}
\newcommand{\eems}{\end{example}}
\newcommand{\els}{\end{lemma}}
\newcommand{\eles}{\end{lemmaex}}
\newcommand{\ers}{\end{remark}}
\newcommand{\ets}{\end{theorem}}
\newcommand{\eds}{\end{definition}}
\newcommand{\eps}{\end{proposition}}

\newcommand{\mybox}{\hfill $\Box$} 
\newcommand{\beq}{\begin{eqnarray}}
\newcommand{\eeq}{\end{eqnarray}}
\newcommand{\beqn}{\begin{eqnarray*}}
\newcommand{\eeqn}{\end{eqnarray*}}
\newcommand{\bi}{\begin{itemize}}
\newcommand{\ei}{\end{itemize}}
\newcommand{\ben}{\begin{enumerate}}
\newcommand{\een}{\end{enumerate}}

\newcommand{\kmo}{\mbox{$k$$-1$}}

\begin{document}

\title{Monotone and near-monotone network structure (part I)}
\author{Eduardo D.\ Sontag%
\footnote{Supported in part by NSf Grants DMS-0504557 and DMS-0614371.}\\
Rutgers University, New Brunswick, NJ, USA\\
\texttt{sontag@math.rutgers.edu}}
\maketitle

\begin{abstract}

This paper provides an expository introduction to monotone and
near-monotone biochemical network structures.
Monotone systems respond in a predictable fashion to perturbations, and have
very robust dynamical characteristics.
This makes them reliable components of more complex networks, and suggests
that natural biological systems may have evolved to be, if not monotone, at
least close to monotone.
In addition, interconnections of monotone systems may be fruitfully analyzed
using tools from control theory.

\end{abstract}

\section{Introduction}

In cells, biochemical networks consisting of proteins, RNA, DNA, metabolites,
and other species, are responsible for control and signaling in development,
regulation, and metabolism, by processing environmental signals, sequencing
internal events such as gene expression, and producing appropriate cellular
responses.
The field of systems molecular biology is largely concerned with the study of
such networks.
Often, as in control theory, biochemical networks are viewed as
interconnections of simpler subsystems.
This paper discusses recent work which makes use of \emph{topology} (graph
structure) as well as \emph{sign} information regarding subsystems and their
interconnection structure in order to infer properties of the complete system.

\begin{figure}[h,t]
 \begin{center}
 \pic{0.2}{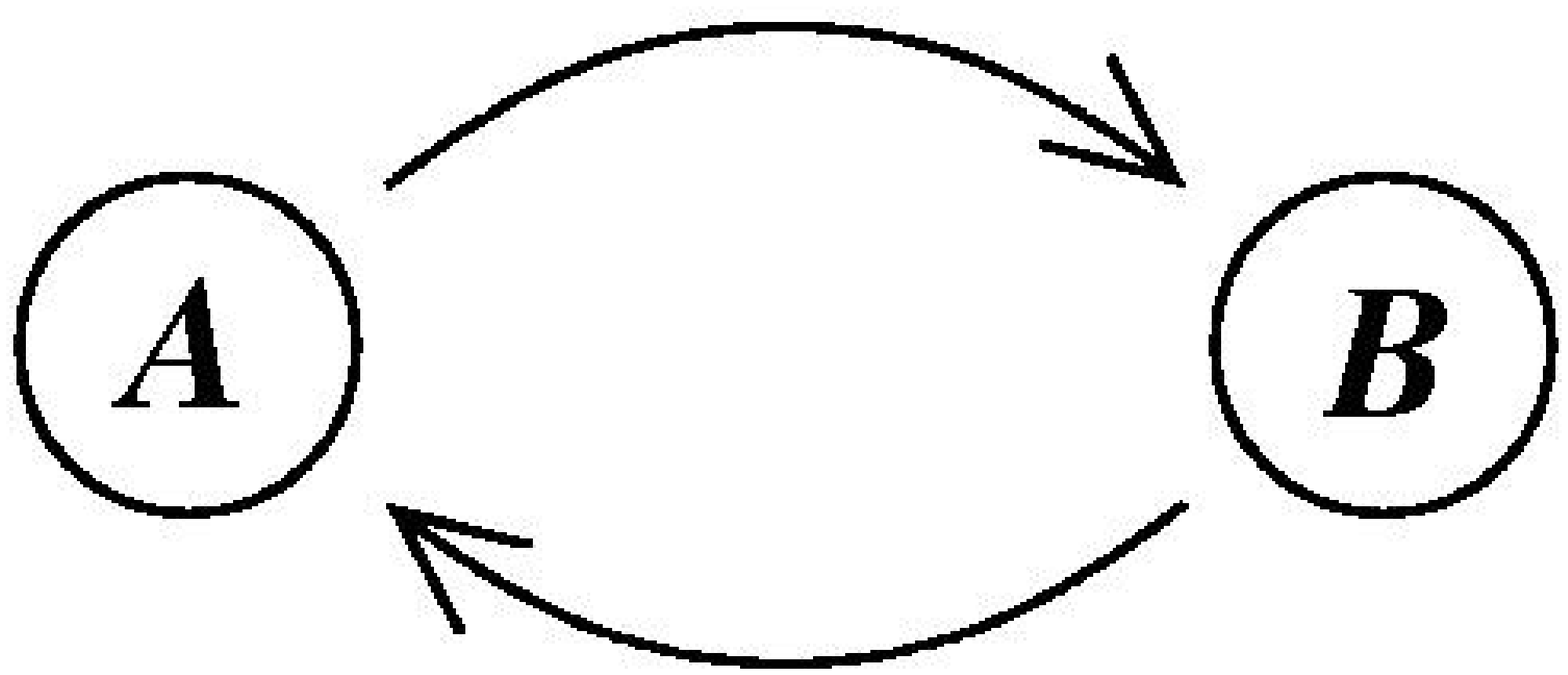}\hskip1cm
 \pic{0.2}{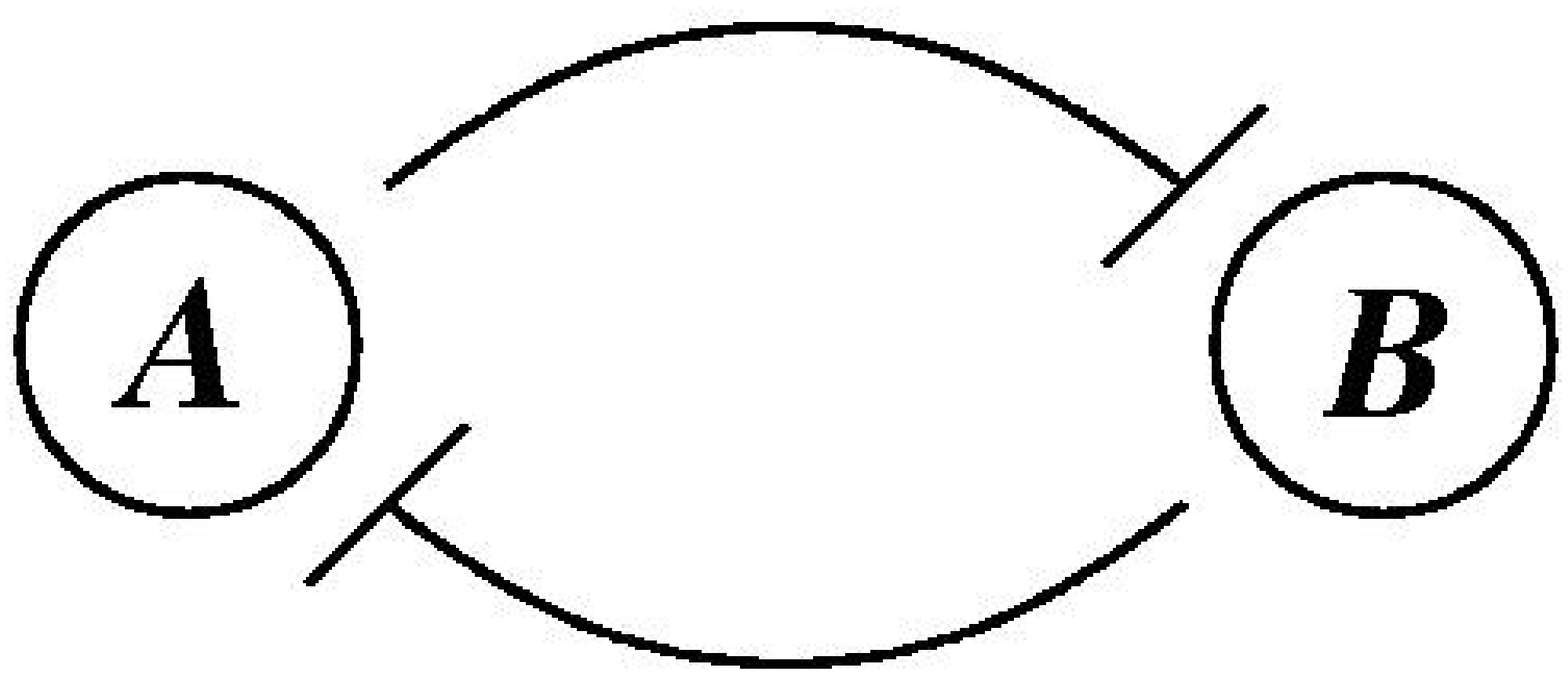}\hskip1cm
 \pic{0.2}{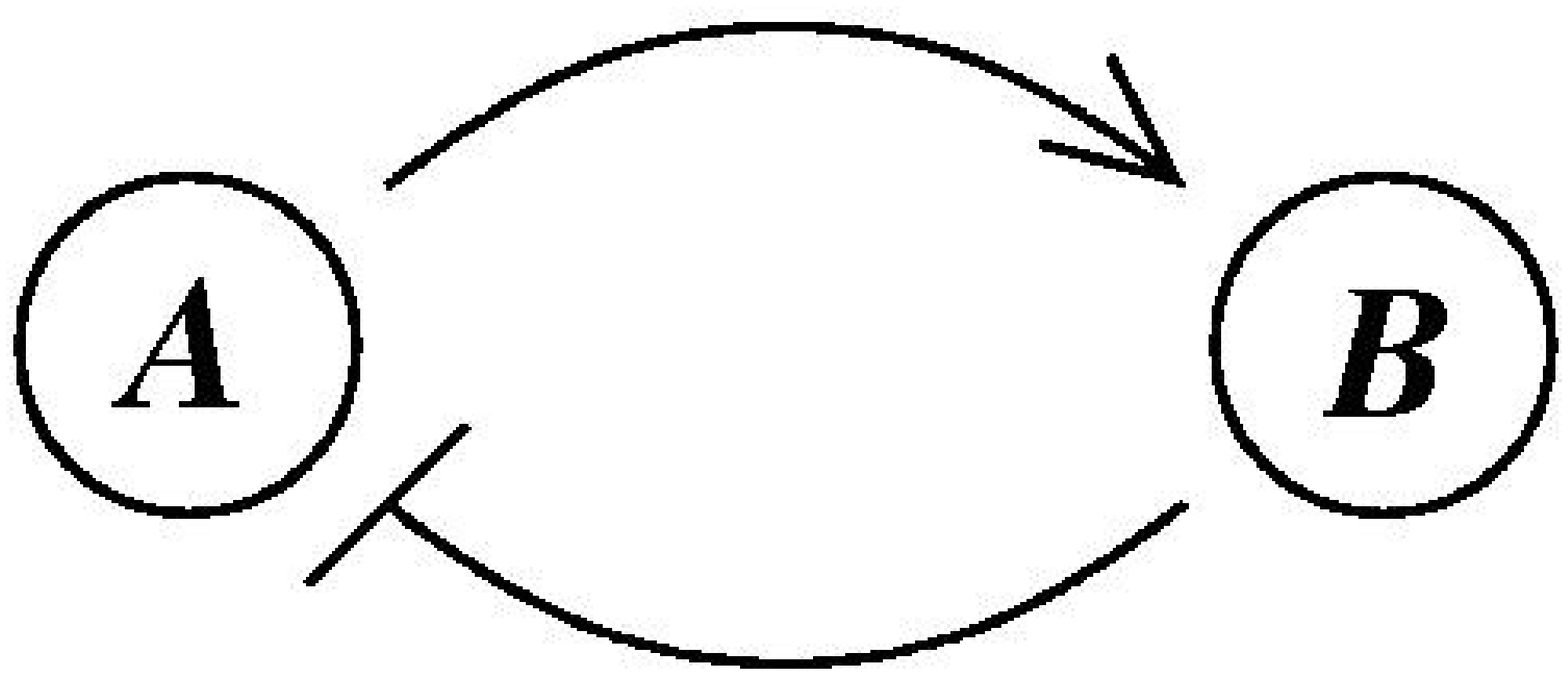}
 \caption{ (a) Mutual activation. $\;\;\;\;\;$
 (b) Mutual inhibition. $\;\;\;$
 (c) Activation-inhibition %
}
 \label{3figs}
 \end{center}
\end{figure}

It is broadly appreciated that behavior is critically dependent on network
topology as well as on the signs (activating or inhibiting) of the
underlying feedforward and feedback interconnections
\cite{novick57,
monod61,
lewis77,
segel84,
deangelis86,
thomas90,
goldbeter96,
keener98,
murray2002,
alon02,
keshet05}%
.
For example, Figures~\ref{3figs}(a-c) show the three possible types of
feedback loops that involve two interacting chemicals,.
A mutual activation configuration is shown in Figure~\ref{3figs}(a):
a positive change in $A$ results in a positive change in $B$, and vice-versa.
Configurations like these are associated to signal amplification and production
of switch-like biochemical responses.
A mutual inhibition configuration is shown in Figure~\ref{3figs}(b):
a positive change in $A$ results in repression of $B$, and repression of $B$
in turn enhances $A$.
Such configurations allow systems to exhibit multiple discrete, alternative
stable steady-states, thus providing a mechanism for memory.
Both (a) and (b) are examples of positive-feedback systems%
~\cite{ptashne92,
plathe95,
cinquin02,
gouze98,
thomas01,
remi03,
monotoneMulti,
pnasangeliferrellsontag04}%
.
On the other hand, activation-inhibition configurations like in
Figure~\ref{3figs}(c) are necessary for the generation of periodic behaviors
such as circadian rhythms or cell cycle oscillations, by themselves or in
combination with multi-stable positive-feedback subsystems, as well as for
adaptation, disturbance rejection, and tight regulation (homeostasis) of
physiological variables
\cite{rapp75,
hastings-tyson-webester-periodic-goodwin-jde77,
tyson-othmer,
thomas81,
keshet05,
goldbeter96,
mct,
kholodenko00,
murray2002,
sha03,
pomerening-ferrell,
monotoneLSU}.
Compared to positive-feedback systems, negative-feedback systems are not
``consistent,'' in a sense to be made precise below but roughly meaning that
different paths between any two nodes should reinforce, rather than contradict,
each other.  For (c), a positive change in $A$ will be resisted by the system
through the feedback loop.
Consistency, or lack thereof,
also plays a role in the behavior of graphs without feedback;
for example~\cite{alon02,alon03,alon3}
deal with the different signal processing capabilities of consistent
(``coherent'') compared to inconsistent feedforward motifs.

A key role in the work to be discussed here will be played by consistent
systems and subsystems.
We will discuss the following points:
\bi
\item
  Interesting and nontrivial conclusions can be drawn from (signed) network
  structure alone.  This structure is associated to purely stoichiometric
  information about the system and ignores fluxes.  Consistency, or close to
  consistency, is an important property in this regard.
\item
  Interpreted as dynamical systems, consistent networks define
  \emph{monotone systems}, which have highly predictable and ordered behavior.
\item
  It is often useful to analyze larger systems by viewing them as
  interconnections of a small number of monotone subsystems.
  This allows one to obtain precise bifurcation diagrams without appeal to
  explicit knowledge of fluxes or of kinetic constants and other parameters,
  using merely ``input/output characteristics'' (steady-state responses or DC
  gains).  The procedure may be viewed as a ``model reduction'' approach in
  which monotone subsystems are viewed as essentially one-dimensional objects.
\item
  The possibility of performing a decomposition into a small number of
  monotone components is closely tied to the question of how ``near'' a
  system is to being monotone.
\item
  We argue that systems that are ``near monotone'' are biologically
  more desirable than systems that are far from being monotone.
\item
  There are indications that biological networks may be much closer to being
  monotone than random networks that have the same numbers of vertices and of
  positive and negative edges.
\ei

\subsubsection*{The need for robust structures and robust analysis tools}

A distinguishing feature of the study of dynamics in cell biology, in
contrast to more established areas of applied mathematics and engineering, is
the high degree of uncertainty inherent in models of cellular
biochemical networks.  This uncertainty is due to environmental fluctuations,
and variability among different cells of the same type, as well as, from a
mathematical analysis perspective, the difficulty of measuring the relevant
model parameters (kinetic constants, cooperativity indices, and many others)
and thus the impossibility of obtaining a precise model.  Thus, it is
imperative to develop tools that are ``robust'' in the sense of being able to
provide useful conclusions based largely upon information regarding the
\emph{qualitative} features of the network, and not the precise values of
parameters or even the forms of reactions.  Of course, this goal is often
not achievable, since dynamical behavior may be subject to phase
transitions (bifurcation phenomena) which are critically dependent on
parameter values. 

\begin{figure}[h,t]
\begin{center}
\setlength{\unitlength}{2700sp}%
\begin{picture}(5775,3232)(1576,-5183)
\thicklines
\put(2701,-3661){\framebox(600,600){}}
\put(4201,-3661){\framebox(600,600){}}
\put(4201,-4861){\framebox(600,600){}}
\put(5701,-4261){\framebox(600,600){}}
\put(3301,-3361){\vector( 1, 0){900}}
\put(4801,-3361){\line( 1, 0){450}}
\put(5251,-3361){\line( 0,-1){450}}
\put(5251,-3811){\vector( 1, 0){450}}
\put(4801,-4561){\line( 1, 0){525}}
\put(5326,-4561){\line( 0, 1){450}}
\put(5326,-4111){\vector( 1, 0){375}}
\put(4501,-3661){\vector( 0,-1){600}}
\put(4201,-4561){\line(-1, 0){1200}}
\put(3001,-4561){\vector( 0, 1){900}}
\put(6001,-3661){\line( 0, 1){825}}
\put(6001,-2836){\line(-1, 0){1500}}
\put(4501,-2836){\vector( 0,-1){225}}
\put(6301,-3961){\vector( 1, 0){900}}
\put(1801,-3361){\vector( 1, 0){900}}
\thicklines
\truered{\put(2401,-5161){\framebox(4200,2550){}}}
{\green\thicklines
\put(4201,-2161){\vector( 1,-1){600}}
\put(4201,-2161){\vector(-1,-2){600}}
\put(6151,-2311){\vector(-1,-1){1575}}
\put(6151,-2311){\vector(-1,-2){675}}%
\black}%
\put(2825,-3436){1}
\put(4325,-3436){2}
\put(4300,-4650){3}
\put(5825,-4036){4}
\put(5200,-2200){outputs of subsystem 2}
\put(2600,-2100){inputs to subsystem 2}
\put(7351,-3900){output}
\put(7351,-4200){of system}
\put(600,-3350){input}
\put(600,-3650){to system}
\end{picture}
\caption{A system composed of four subsystems}
\label{fig-systems}
\end{center}
\end{figure}
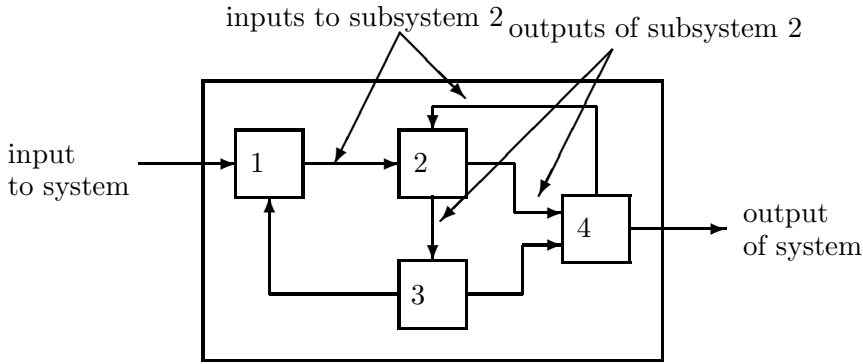

Nevertheless, and surprisingly, research by many, notably by
Clarke~\cite{clarke},
Horn and Jackson~\cite{hornjackson1,hornjackson2}, and
Feinberg~\cite{feinberg0,feinberg1,feinberg2}, relying upon
\emph{complex balancing and deficiency theory}, has resulted in the
identification of rich classes of chemical network structures for which such
robust analysis is indeed possible.
A different direction of work, pioneered by Hirsch and Smith~\cite{smith,Hirsch-Smith},
relies upon the theory of \emph{monotone systems}, and has a 
similar goal of drawing conclusions about dynamical behavior based only upon
structure.  This direction has been enriched substantially by the introduction
of \emph{monotone systems with inputs and outputs}: as standard in control
theory~\cite{mct}, one extends the notion of monotone system so as to
incorporate input and output channels~\cite{monotoneTAC}.
Once inputs and outputs are introduced, one can study interconnections
of systems (Figure~\ref{fig-systems}),
and ask what special properties hold if the subsystems are
monotone~\cite{monotoneTAC,JMathChemMonotone06,pnasangeliferrellsontag04}. 

\section{Consistent Graphs, Monotone Systems, and Near-Monotonicity}

We now introduce the basic notions of monotonicity and consistency.
The present section deals exclusively with graph-theoretic information,
which is derived from stoichiometric constraints.
Complementary to this analysis,
bifurcation phenomena can be sometimes analyzed using a combination of
these graphical techniques together with information on steady-state 
gains; that subject is discussed in Part II.
The discussion is informal; in Part II has
more rigorous mathematical statements, presented in the
more general context of systems with external inputs and outputs.

The systems considered here are described by the evolution of
states, which are time-dependent vectors
$x(t)=(x_1(t),\ldots ,x_n(t))$ whose components $x_i$ represent concentrations of
chemical species such as proteins, mRNA, or metabolites.
In autonomous
differential equation (``continuous-time'') models, one specifies the rate
of change of each variable, at any given time, as a function of the
concentrations of all the variables at that time:
\beqn
\frac{dx_1}{dt}(t) &=& f_1(x_1(t),x_2(t),\ldots ,x_n(t))\\
\frac{dx_2}{dt}(t) &=& f_2(x_1(t),x_2(t),\ldots ,x_n(t))\\
&\vdots&\\
\frac{dx_n}{dt}(t) &=& f_n(x_1(t),x_2(t),\ldots ,x_n(t))\,,
\eeqn
or just $dx/dt = f(x)$, where $f$ is the vector function with components $f_i$.
We assume that the coordinates $x_i$ of the state of the system can be
arbitrary non-negative numbers.  (Constraints among variables can be imposed as
well, but several aspects of the theory are more subtle in that case.)
Often, one starts from a differential equation system written in the
following form:
\[
\frac{dx}{dt}(t) = \Gamma R(x),
\]
where $R(x)$ is a $q$-dimensional \emph{vector of reactions} and
$\Gamma $ is an $n\times q$ matrix, called the \emph{stoichiometry matrix}, and
either one studies this system directly, or one studies a smaller set of
differential equations $dx/dt = f(x)$ obtained by eliminating variables
through the use of conserved stoichiometric quantities.

We will mostly discuss differential equation models, but will also make
remarks concerning difference equation (``discrete time'') models.
The dynamics of these are described by rules
that specify the state at some future time $t=t_{k+1}$ as a function of the
state of the system at the present time $t_k$. 
Thus, the $i$th coordinate evolves according to an update rule:
\[
x_i(t_{k+1})= f_i(x_1(t_k),x_2(t_k),\ldots ,x_n(t_k))
\]
instead of being described by a differential equation.  Usually, $t_k=k\Delta $,
where $\Delta $ is a uniform inter-sample time.
One may associate a difference equation to any given differential equation,
through the rule that the vector $x(t_{k+1})$ should equal the solution of the
differential equation when starting at state $x(t_k)$.  However, not every
difference equation arises from a differential equation in this manner.
Difference equations may be more natural when studying processes in which
measurements are made at discrete times, or they might provide a 
macroscopic model of an underlying stochastic process taking place at a faster
time scale.

One may also study more complicated descriptions of dynamics that those given
by ordinary differential and difference equations; many of
the results that we discuss
here have close analogs that apply to more general classes of (deterministic)
dynamical systems, including reaction-diffusion partial differential
equations, which are used for space-dependent problems with slow diffusion and
no mixing, delay-differential systems, which help model delays due to
transport and other cellular phenomena in which concentrations of one species
only affect others after a time interval, and integro-differential equations
\cite{smith,Hirsch-Smith,sysbio04,enciso_smith_sontagJDE06}.
In a different direction, one may consider systems with external inputs and
outputs
\cite{monotoneTAC}.

\subsubsection*{The graph associated to a system}

There are at least two types of graphs that can be naturally associated to a
given biochemical network.  One type, sometimes called the
\emph{species-reaction graph}, is a bipartite graph with nodes for reactions
(fluxes) and species, which leads to useful analysis techniques based on Petri
net theory and graph theory
\cite{feinberg3,reddy,schuster,craciun1,craciun2,translation-invariance,persistencePetri,06cdc_chemical}.
We will not discuss species-reaction graphs here.
A second type of graph, which we will discuss, is the \emph{species graph} $G$.
It has $n$ nodes (or ``vertices''),
which we denote by $v_1,\ldots ,v_n$, one node for each species.
No edge is drawn from node $v_j$ to node $v_i$ if the partial derivative
$\frac{\partial f_i}{\partial x_j}(x)$ vanishes identically, meaning that
there is no direct effect of the $j$th species upon the $i$th species.
If this derivative is not identically zero, then there are three possibilities:
(1) it is $\geq 0$ for all $x$, 
(2) it is $\leq 0$ for all $x$,
or
(3) it changes sign depending on the particular entries of the concentration
vector $x$.
In the first case (activation), we draw an edge labeled $+$, $+1$, or just an
arrow $\rightarrow $. 
In the second case (repression or inhibition), we draw an edge labeled $-$,
$-1$, or use the symbol $\dashv$.
In the third case, when the sign is ambiguous, we draw both an activating
and an inhibiting edge from node $v_j$ to node $v_i$.

For continuous-time systems, no self-edges (edges from a node $v_i$ to 
itself) are included in the graph $G$, whatever the sign of
the diagonal entry $\partial f_i/\partial x_i$ of the Jacobian.
For discrete-time systems, on the other hand,
self-edges are included (we later discuss the reason for these different
definitions for differential and difference equations).

\begin{figure}[h,t]
 \begin{center}
 \pic{0.45}{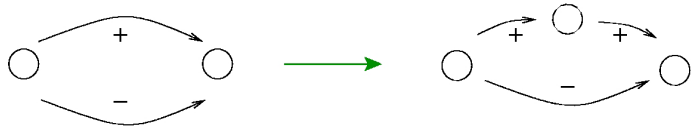}
 \caption{Replacing direct inconsistent effects by adding a node}
 \label{adding_node_direct_effect}
 \end{center}
\end{figure}

When working with graphs, it is more convenient (though not strictly
necessary) to consider only graphs $G$ that have no multiple edges from one
node to another (third case above).
One may always assume that $G$ has this property, by means of the
following trick: whenever there are two edges, we replace one of them by an
indirect link involving a new node; see \Fig{adding_node_direct_effect}.
Introducing such additional nodes if required, we will suppose from now on
that no multiple edges exist.

Although adding new edges as explained above is a purely formal construction
with graphs, it may be explained biologically as follows.  Often, ambiguous
signs in Jacobians reflect heterogeneous mechanisms.  For example,
take the case where protein A enhances the transcription rate of gene B if
present at high concentrations, but represses B if its concentration is
lower than some threshold.  Further study of the chemical mechanism might
well reveal the existence of, for example, a homodimer that is responsible for
this ambiguous effect.  Mathematically, the rate of transcription of B might be
given algebraically by the formula $k_2a^2-k_1a$, where $a$ denotes the
concentration of A.  Introducing a new species C to represent the
homodimer, we may rewrite this rate as $k_2c-k_1a$, where $c$ is the
concentration of C, plus an new equation like $dc/dt = k_3a^2-k_4c$
representing the formation of the dimer and its degradation. 
This is exactly the situation in \Fig{adding_node_direct_effect}.

\subsubsection*{Spin assignments and consistency}

A \emph{spin assignment} $\Sigma $ for the graph $G$ is an assignment,
to each node $v_i$, of a number $\sigma _i$ equal to ``$+1$'' or ``$-1$'' (a
``spin,'' to borrow from statistical mechanics terminology).
In graphical depictions, we draw up-arrows or down-arrows to indicate spins.
If there is an edge from node $v_j$ to node $v_i$, with label
$J_{ij}\in \{\pm1\}$,
we say that this edge is \emph{consistent with the spin assignment $\Sigma $}
provided that:
\[
J_{ij}\sigma _i\sigma _j=1
\]
which is the same as saying that $J_{ij}=\sigma _i\sigma _j$, or that
$\sigma _i=J_{ij}\sigma _j$.
An equivalent formalism is that in which edges are labeled by ``$0$''
or ``$1$,'' instead of $1$ and $-1$ respectively, and edge labels
$J_{ij}$ belong to the set $\{0,1\}$, in which case
consistency is the property 
that $J_{ij}\oplus\sigma _i\oplus\sigma _j=0$ (sum modulo two).

We will say that \emph{$\Sigma $ is a consistent spin assignment for
the graph $G$}
(or simply that $G$ is consistent)
if every edge of $G$ is consistent with $\Sigma $.
In other words, for any pair of vertices $v_i$ and $v_j$, if there is a
positive edge from node $v_j$ to node $v_i$, then $v_j$ and $v_i$ must have
the same spin, and if there is a negative edge connecting $v_j$ to $v_i$, then
$v_j$ and $v_i$ must have opposite spins.
(If there is no edge from $v_j$ to $v_i$, this requirement imposes no
restriction on their spins.)

In order to decide whether a graph admits any consistent spin assignment, it is
not necessary to actually test all the possible $2^n$ spin assignments.
It is very easy to prove that there is a consistent assignment if and only if 
\emph{every undirected loop in the graph $G$ has a net positive sign}, that
is to say, an even number, possibly zero, of negative arrows.  Equivalently,
any two (undirected) paths between two nodes must have the same net sign.
By undirected loops or paths, we mean that one is allowed to transverse an
edge either forward or backward.
A proof of this condition is as follows.
If a consistent assignment exists, then, for any undirected loop
$v_{i_1},\ldots ,v_{i_k}=v_{i_1}$ starting from and ending at the node $v_{i_1}$,
inductively one has that:
\[
\sigma _{i_1} \;= \; Q_{i_1,i_{k-1}} \, Q_{i_{k-1},i_{k-2}} 
        \,\ldots , \, Q_{i_{2},i_{1}}\,\sigma _{i_1}
\]
where $Q_{ij}=J_{ij}$ if we are transversing the edge from $v_j$ to $v_i$, or
$Q_{ij}=J_{ji}$ if we are transversing backward the edge from $v_j$ to $v_i$.
This implies (divide by $\sigma _{i_1}$) that the product of the edge signs is
positive.
Conversely, if any two paths between nodes have the same parity, and the graph
is connected, pick node $v_1$ and label it ``$+$'' and then assign to every
other node $v_i$ the parity of a path connecting $v_1$ and $v_i$.
(If the graph is not connected, do this construction on each component
separately.) 

This positive-loop property, in turn, can be checked with a fast dynamic
programming-like algorithm.  For connected graphs, there can be at most two
consistent assignments, each of which is the reverse (flip every spin) of the
other.

\subsubsection*{Monotone systems}

A dynamical system is said to be \emph{monotone} if there exists at least one
consistent spin assignment for its associated graph $G$.
Monotone systems~\cite{Hirsch,Hirsch2,smith} were introduced by Hirsch,
and constitute a class of dynamical systems for which a rich theory exists.
(To be precise, we have only defined the subclass of systems that are
\emph{monotone with respect to some orthant order}.  The notion of monotonicity
can be defined with respect to more general orders.)

\subsubsection*{Consistent response to perturbations}

\begin{figure}[h,t]
 \begin{center}
 \pic{0.2}{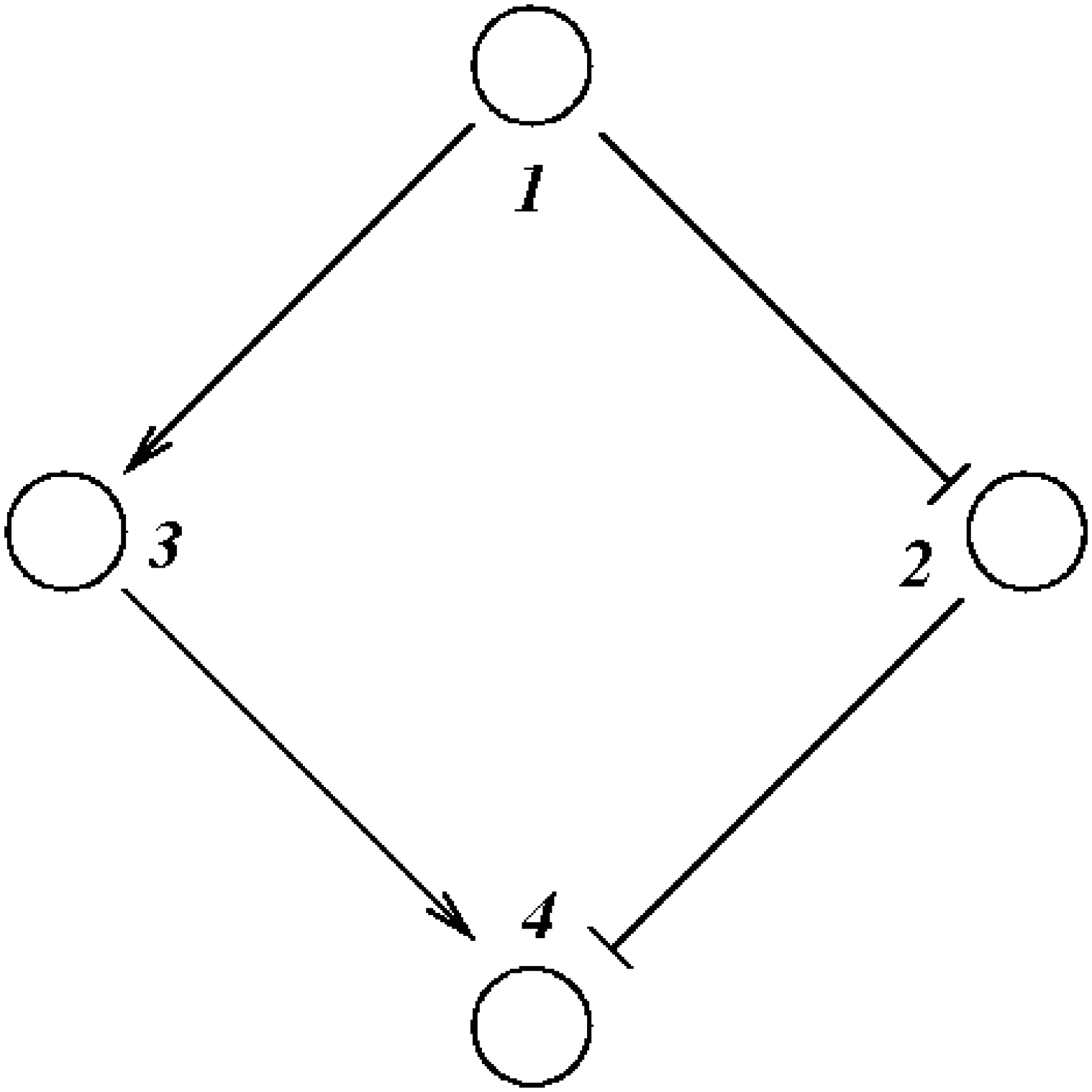}\hskip0.3cm
 \pic{0.2}{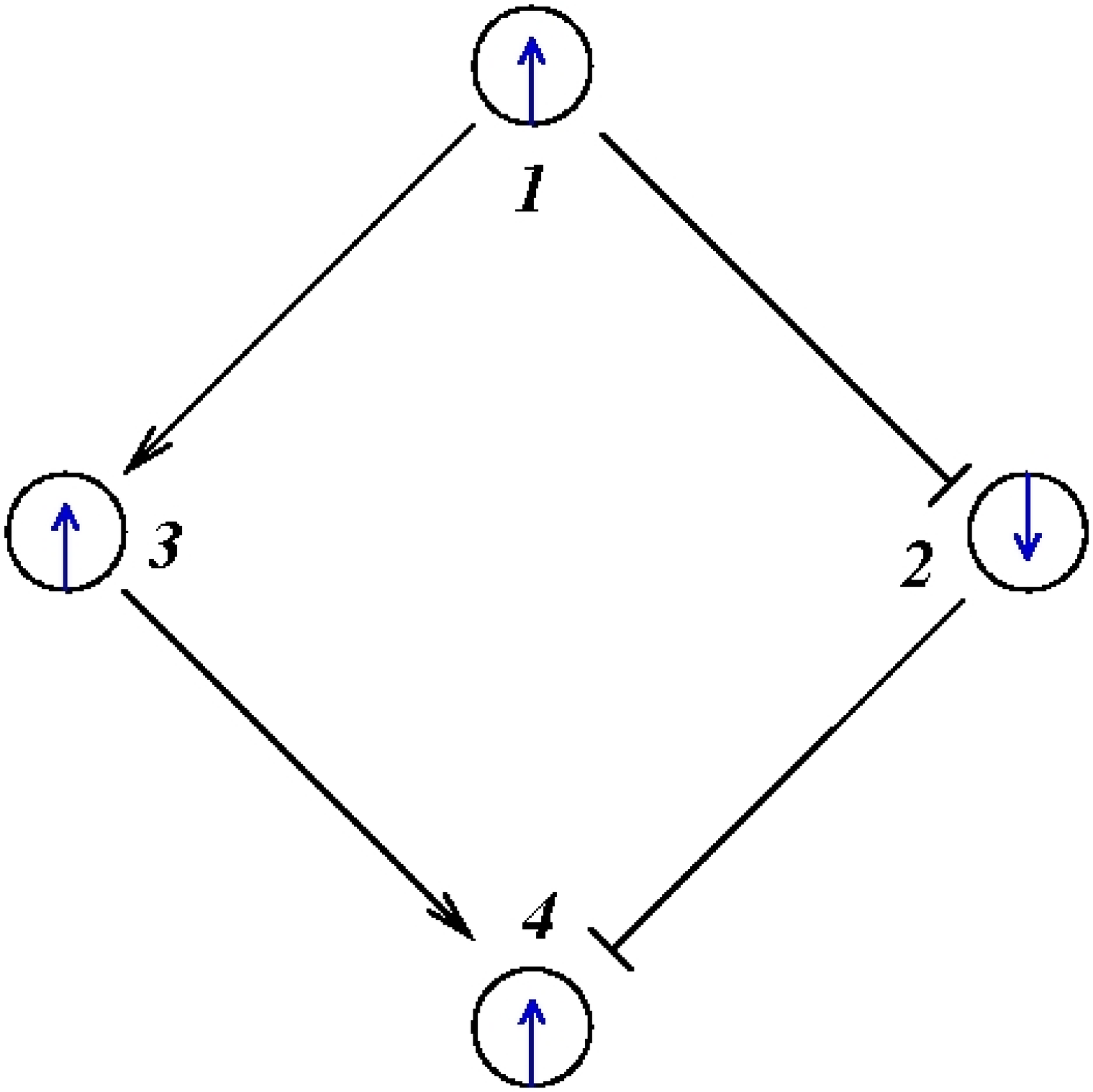}\hskip0.3cm
 \pic{0.2}{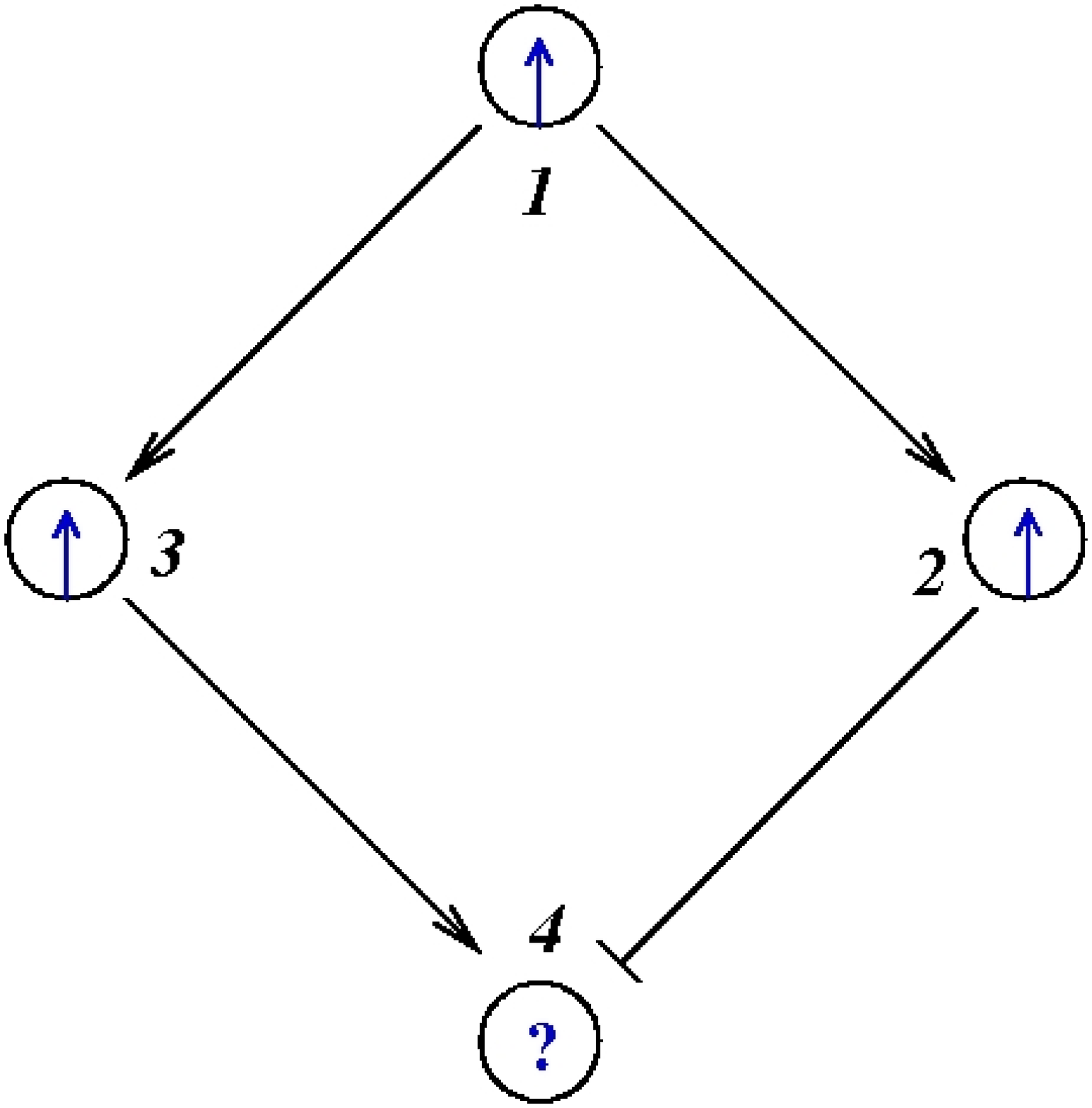}\hskip0.3cm
 \pic{0.2}{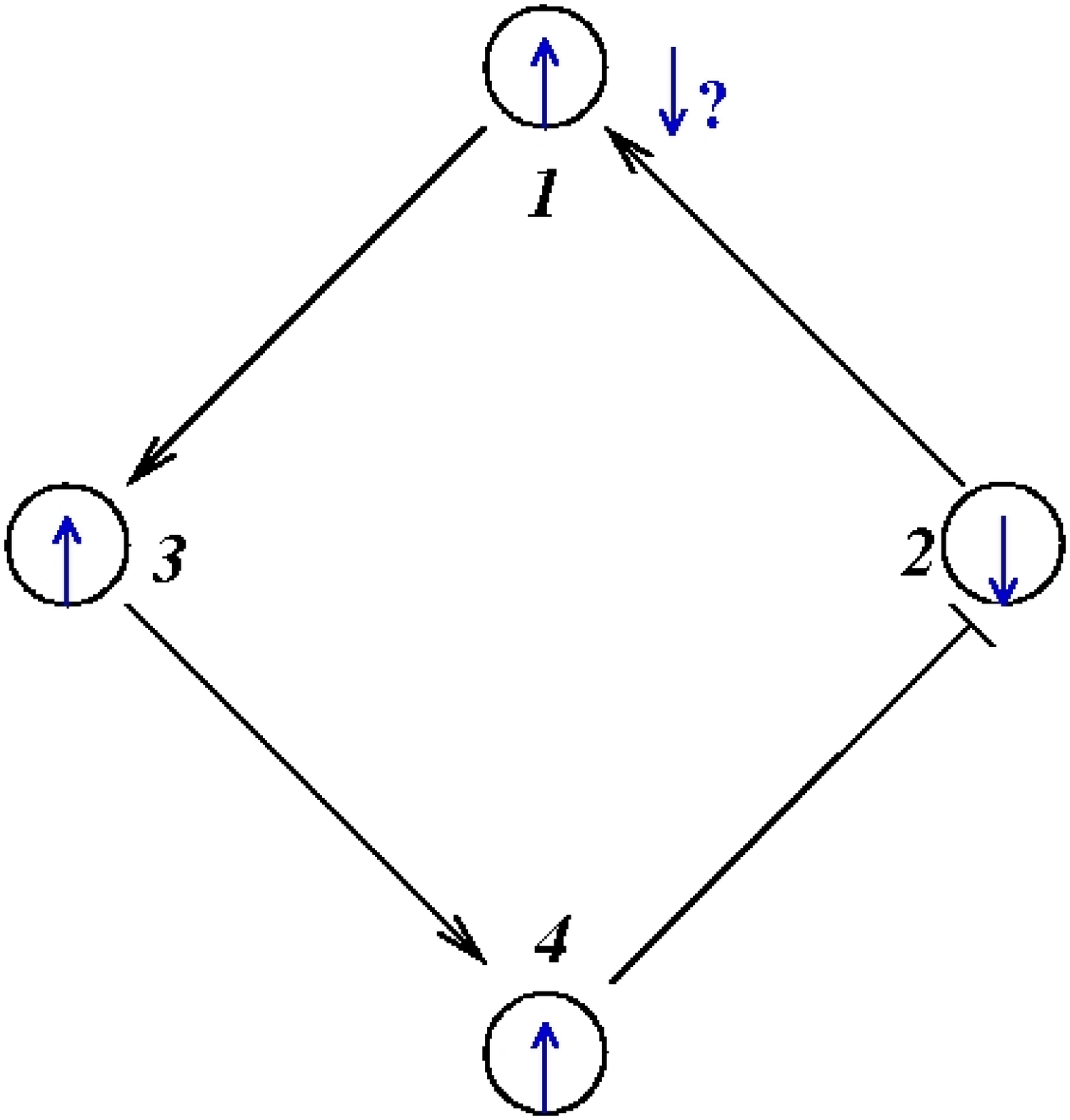}
 \caption{(a) and (b) graph and consistent assignment, (c) and (d) no possible consistent assignments
}
 \label{consistent_fig}
 \end{center}
\end{figure}

Monotonicity reflects the fact that a system responds consistently to
perturbations on its components.  Let us now discuss this property in informal
terms.
We view the nodes of the graph shown 
in Figure~\ref{consistent_fig}(a)
as corresponding to variables in the system, which quantify the concentrations
of chemical species such as activated receptors, proteins, transcription
factors, and so forth.  
Suppose that a perturbation, for example due to the external
activation of a receptor represented by node 1, 
instantaneously increases the value of the concentration of this species. 
We represent this increase by an up-arrow inserted into that node, as in
Figure~\ref{consistent_fig}(b).
The effect on the other nodes is then completely predictable from the
graph.  The species associated to node 2 will decrease, because of the
inhibiting character of the connection from 1 to 2, and the species associated
to node 3 will increase (activating effect).  
Where monotonicity plays a role is in insuring that the concentration of
the species corresponding to node 4 will also increase.
It increases both because it is activated by 3, which has increased, and 
because it is inhibited by 2, so that less of 2 implies a smaller inhibition
effect. 
Algebraically, the following expression involving partial derivatives:
\[
\frac{\partial f_4}{\partial x_3} \frac{\partial f_3}{\partial x_1}
\;+\;
\frac{\partial f_4}{\partial x_2} \frac{\partial f_2}{\partial x_1} 
\]
(where $f_i$ gives the rate of change of the $i$th species, in the
differential equation model)
is guaranteed to be positive, since it is a sum of positive terms:
$(+)(+) + (-)(-)$.
Intuitively, the expression measures the sensitivity of the rate of change
$dx_4/dt$ of the concentration of 4 with respect to perturbations in 1,
with the two terms giving the contributions for each of the two alternative
paths from node 1 to node 4. 
\emph{This unambiguous global effect holds true regardless of the actual
values of parameters such as kinetic constants, and even the algebraic forms
of reactions, and depends only on the signs of the entries of the Jacobian of
$f$.}
Observe that the arrows shown in Figure~\ref{consistent_fig}(b) provide
a consistent spin-assignment for the graph, so the system is monotone.

In contrast, consider next the graph in Figure~\ref{consistent_fig}(c),
where the edge from 1 to 2 is now positive.
There are two paths from node 1 to node 4, one of which (through 3) is
positive and the other of which (through 2) is negative.  Equivalently,
the undirected loop $1,3,4,2,1$ (``undirected'' because the
last two edges are transversed backward) has a net negative parity.
Therefore, the loop test for consistency fails, so that there is no possible
consistent spin-assignment for this graph, and therefore the corresponding
dynamical system is not monotone. 
Reflecting this fact, the net effect of an increase in node 1 is ambiguous.
\emph{It is impossible to conclude from the graphical information alone}
whether node 4 will be repressed (because of the path through 2) or
activated (because of the path through 3).  There is no way to
resolve this ambiguity unless equations and precise parameter values are
assigned to the arrows.

To take a concrete example, suppose that the equations for the system are as
follows:
\[
\frac{dx_1}{dt}=0\quad\quad\quad
\frac{dx_2}{dt}=x_1\quad\quad\quad
\frac{dx_3}{dt}=x_1\quad\quad\quad
\frac{dx_4}{dt}=x_4(k_3x_3-k_2x_2)\,,
\]
where the reaction constants $k_2$ and $k_3$ are two positive numbers.
The initial conditions are taken to be $x_1(0)=x_4(0)=1$,
and $x_2(0)=x_3(0)=0$, and we ask how the solution $x_4(t)$ will
change when the initial value $x_1(0)$ is perturbed.
With $x_1(0)=1$, the solution is $x_4(t)=e^{\alpha t^2/2}$, where $\alpha =k_3-k_2$.
On the other hand, if $x_1(0)$ is perturbed to a larger value, let us say
$x_1(0)=2$,
then $x_4(t)=e^{\alpha t^2}$.  This new value of $x_4(t)$ is larger than
the original unperturbed value $e^{\alpha t^2/2}$ provided that $\alpha >0$,
but it is smaller than it if, instead, $\alpha <0$.  In other words, the sign of
the sensitivity of $x_4$ to a perturbation on $x_1$ cannot be predicted from
knowledge of the graph alone, but it depends on whether $k_2<k_3$ or $k_2>k_3$.
Compare this with the monotone case, as in Figure~\ref{consistent_fig}(a).
A concrete example is obtained if we modify the $x_2$ equation to
$\fractextnp{dx_2}{dt}=\fractextnp{1}{(1+x_1)}$.
Now the solutions are $x_4(t)=e^{\beta _1t^2}$
and $x_4(t)=e^{\beta _2t^2}$ respectively, with $\beta _1=k_3/2-k_2/4$
and $\beta _2=k_3-k_2/6$, so we are guaranteed
that $x_4$ is larger in the perturbed case, a conclusion that holds true no
matter what are the numerical values of the (positive) constants $k_i$.

The uncertainty associated to a graph like the one in
Figure~\ref{consistent_fig}(c) might be undesirable in natural systems.  Cells
of the same type differ in concentrations of ATP, enzymes, and other
chemicals, and this affects the values of model parameters, so two cells of
the same type may well react differently to the same ``stimulus''
(increase in concentration of chemical 1).
While such epigenetic diversity is sometimes desirable, it makes behavior
less predictable and robust.
From an evolutionary viewpoint, a ``change in wiring'' such as replacing the
negative edge from 1 to 2 by a positive one (or, instead, perhaps introducing
an additional inconsistent edge) could lead to unpredictable effects, and so
the fitness of such a mutation may be harder to evaluate.
In a monotone system, in contrast, a stimulus applied to a component is
propagated in an unambiguous manner throughout the circuit, promoting a
predictably consistent increase or consistent decrease in the concentrations
of all other components.

Similarly, consistency also applies to feedback loops.  For example,
consider the graph shown in 
Figure~\ref{consistent_fig}(d).
The negative feedback given by the inconsistent path $1,3,4,2,1$ means that
the instantaneous effect of an up-perturbation of node 1 feeds back into a
negative effect on node 1, while a down-perturbation feeds back as a positive
effect.  In other words, the feedback loop acts against the perturbation.

Of course, negative feedback as well as inconsistent feedforward circuits are
important components of biomolecular networks, playing a major role in
homeostasis and in signal detection.  The point being made here is that
inconsistent networks may require a more delicate tuning in order to perform
their functions.

In rigorous mathematical terms, this predictability property can be formulated
as \emph{Kamke's Theorem}.
Suppose that $\Sigma =\{\sigma _i,i=1,\ldots ,n\}$ is a consistent spin assignment for the
system graph $G$.  
Let $x(t)$ be any solution of $dx/dt=f(x)$.
We wish to study how the solution $z(t)$ arising from a perturbed initial
condition $z(0)=x(0)+\Delta $ compares to the solution $x(t)$.
Specifically, suppose that a positive perturbation is performed at time $t=0$
on the $i$th coordinate, for some index $i\in \{1,\ldots ,n\}$: $z_i(0)>x_i(0)$
and $z_j(0)=x_j(0)$ for all $j\not= i$.
For concreteness, let us assume that the perturbed node $i$ has been labeled
by $\sigma _i=+1$.
Then, Kamke's Theorem says the following: for each node that has the same
parity (i.e., each index $j$ such that $\sigma _j=+1$), and for every future time
$t$, $z_j(t)\geq x_j(t)$.  Similarly, for each node with opposite parity
($\sigma _j=-1$), and for every time $t$, $z_j(t)\leq x_j(t)$.  (Moreover, one or more
of these inequalities must be strict.)
This is the precise sense in which an up-perturbation of the species
represented by node $v_i$ unambiguously propagates into up- or down-behavior
of all the other species.
See~\cite{smith} for a proof, and see~\cite{monotoneTAC} for
generalizations to systems with external input and output channels.

For difference equations (discrete time systems), once that that self-loops
have been included in the graph $G$ and the definition of consistency, Kamke's
theorem also holds; in this case the proof is easy, by induction on time
steps.

\begin{figure}[h,t]
 \begin{center}
 \pic{0.2}{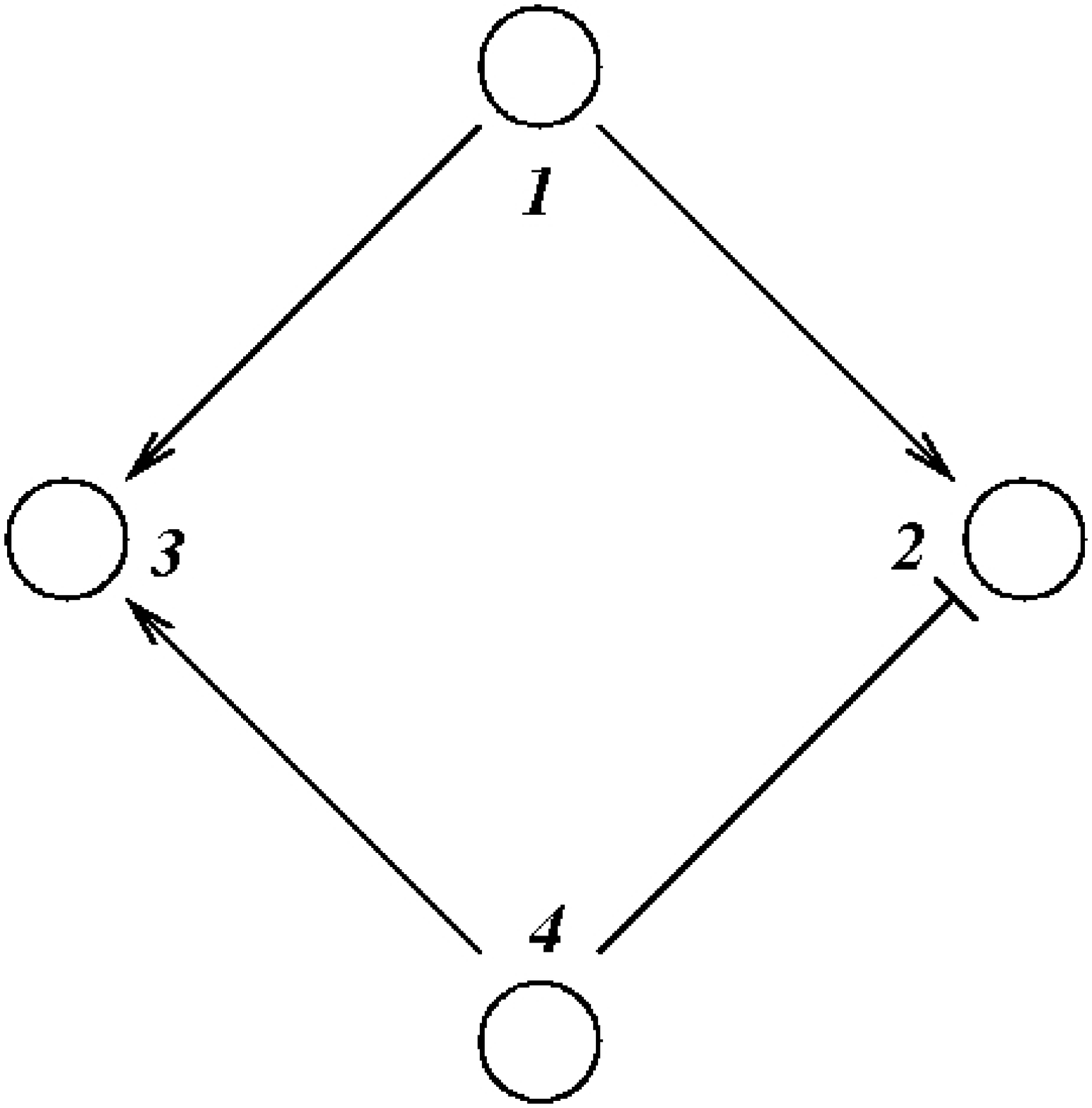}\hskip0.6cm
 \pic{0.2}{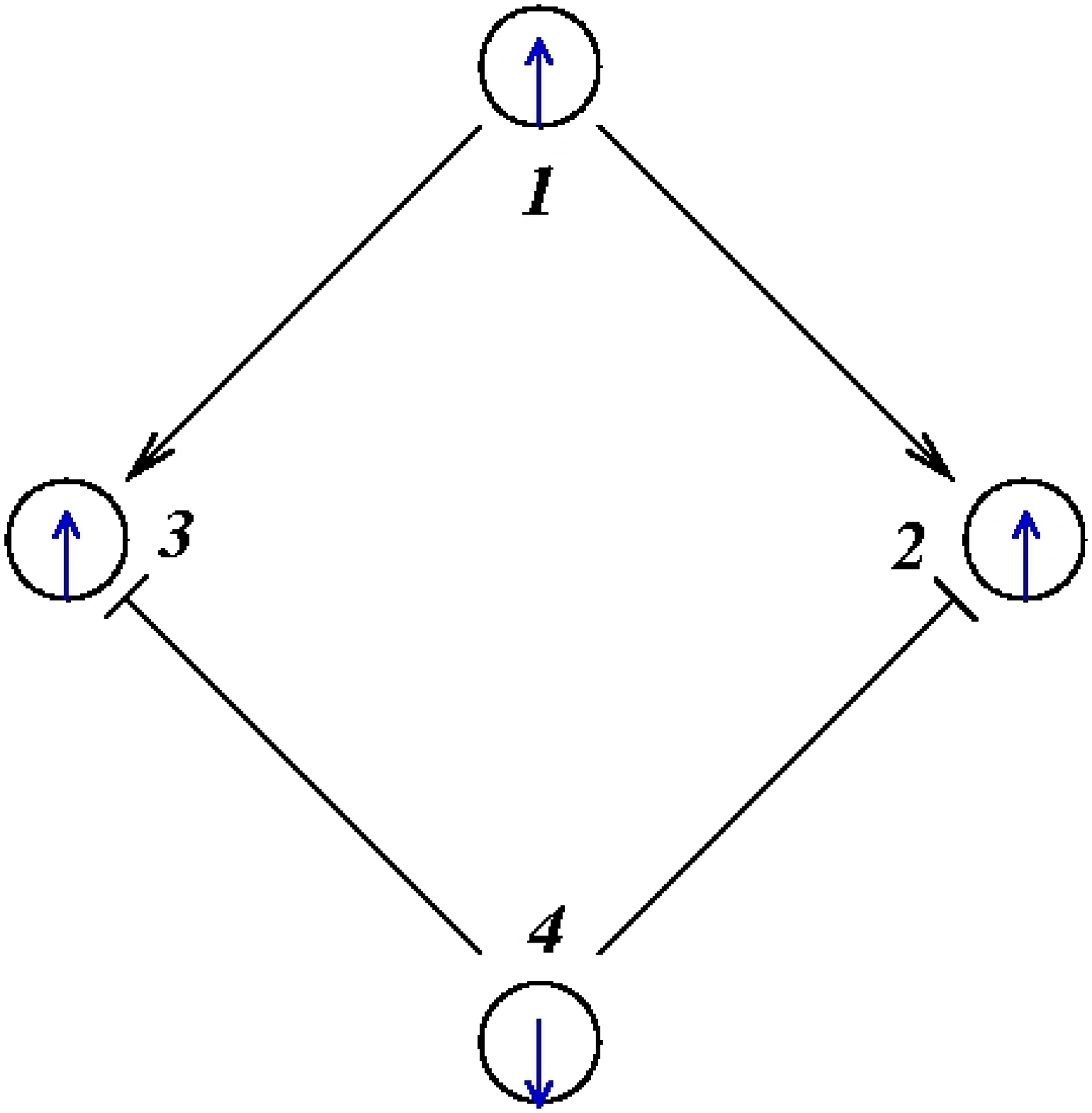}\hskip0.6cm
 \pic{0.28}{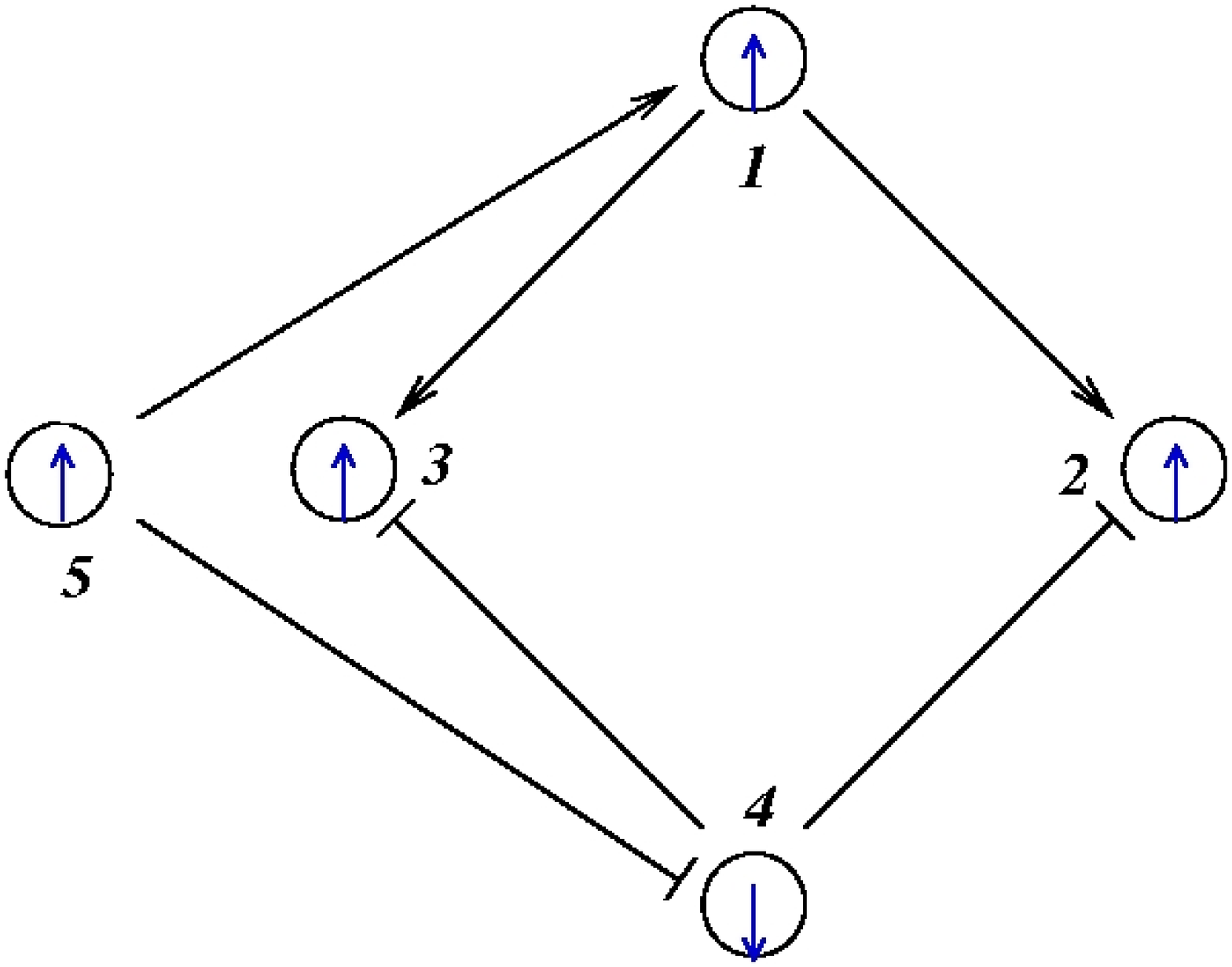}
 \caption{(a) inconsistent, (b) consistent, (c) adding node to consistent network}
 \label{consistent_fig_rev}
 \end{center}
\end{figure}

Consistent graphs can be embedded into larger
consistent ones, but inconsistent ones cannot.  For example,
consider the graph shown in
Figure~\ref{consistent_fig_rev}(a).
This graph admits no consistent spin assignment since the
undirected loop $1,3,4,2,1$ has a net negative parity.
Thus, there cannot be any consistent graph that includes this graph as a
subgraph.  Compare this with the graph shown in
Figure~\ref{consistent_fig_rev}(b).
Consistency of this graph may well represent consistency of a larger graph
which involves a yet-undiscovered species, such as node 5 in
Figure~\ref{consistent_fig_rev}(c).
Alternatively, and from an ``incremental design'' viewpoint, this graph being
consistent makes it possible to consistently add node 5 in the future.

\subsubsection*{Removing the smallest number of edges so as to achieve consistency}

Let us call the \emph{consistency deficit (CD)} of a graph $G$ the smallest
possible number of edges that should be removed from $G$ in order that there
remains a consistent graph, and, correspondingly, a monotone system.

\begin{figure}[h,t]
 \begin{center}
 \pic{0.2}{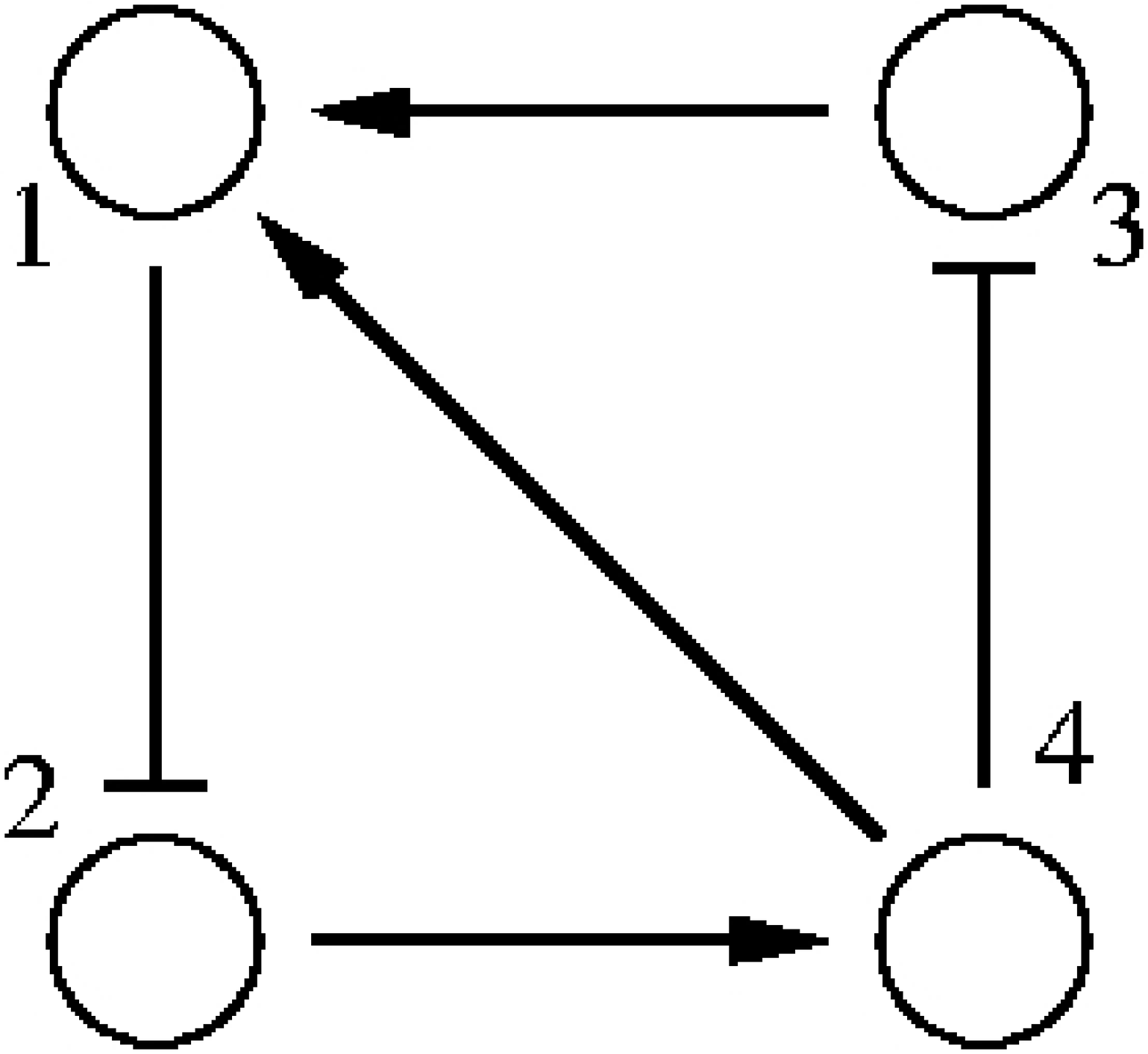}\hskip0.6cm
 \pic{0.2}{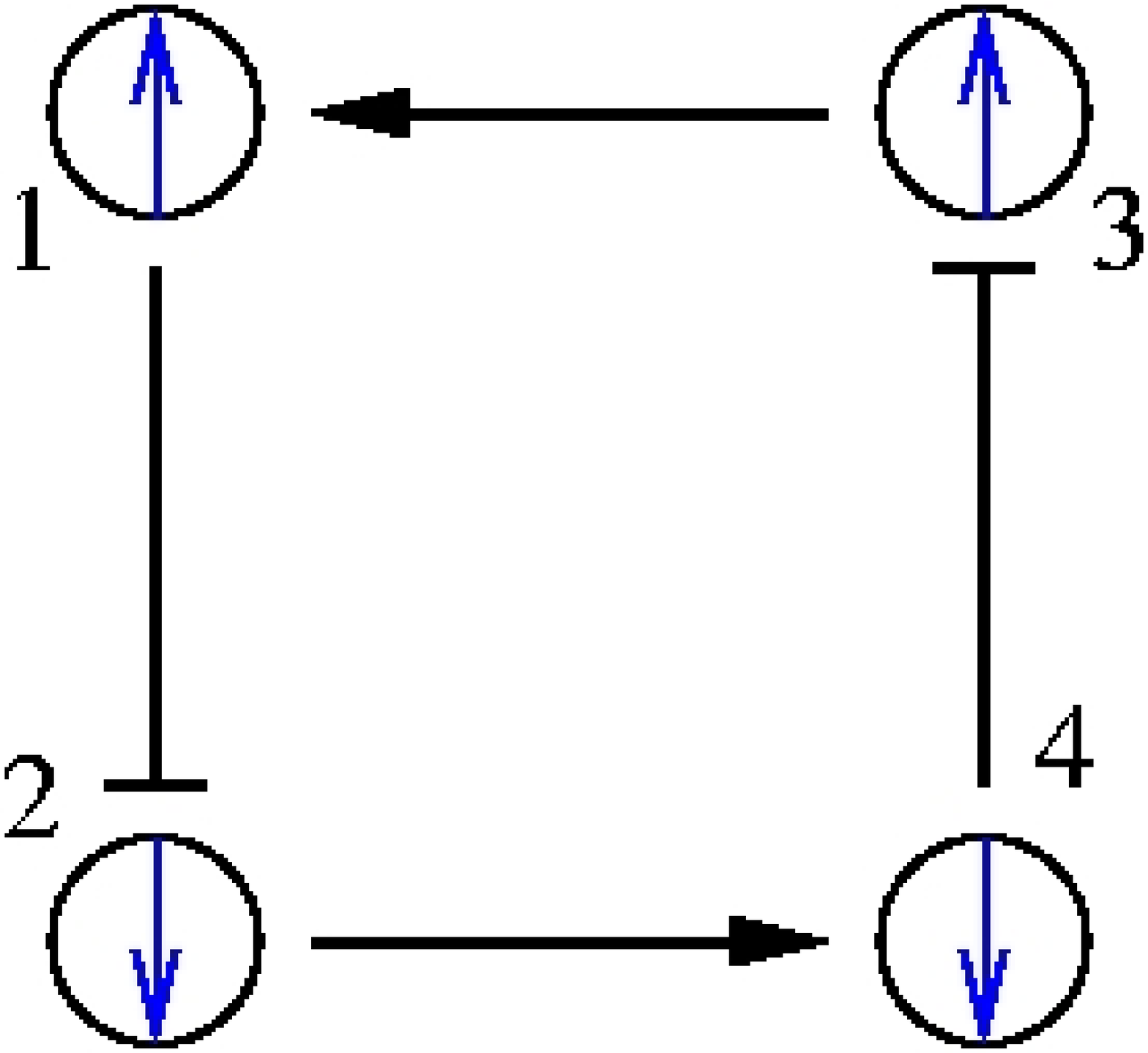}\hskip0.6cm
 \pic{0.2}{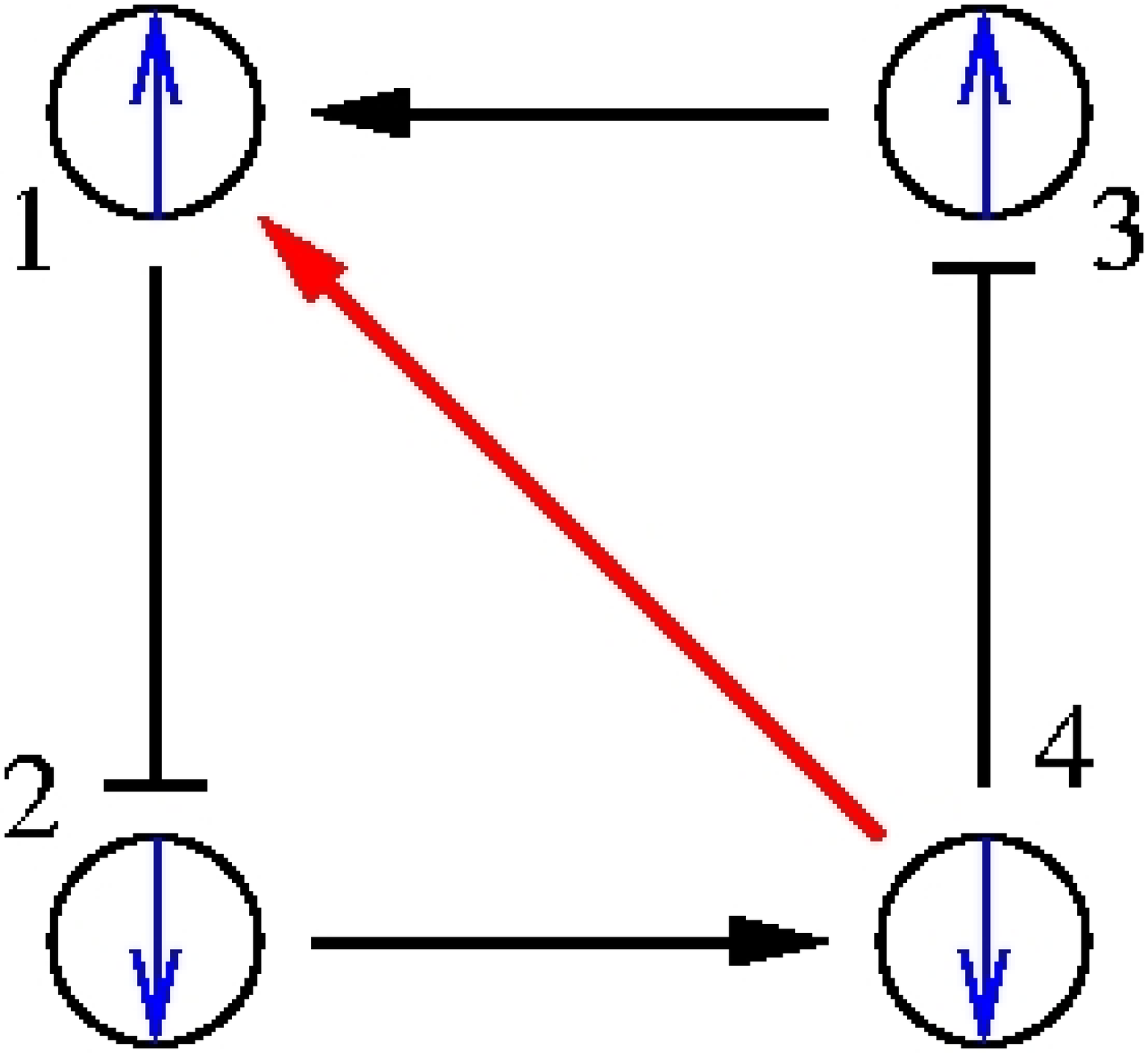}
 \caption{(a) inconsistent graph, (b) consistent subgraph, (c) one
 inconsistent edge}
 \label{5-node-figure}
 \end{center}
\end{figure}

As an example, take the graph shown in Figure~\ref{5-node-figure}(a).
For this graph, it suffices to remove just one edge, the diagonal positive
one, so the CD is 1.
(In this example, the solution is unique, in that no other single other edge
would suffice, but for other graphs there are typically several alternative
ways to achieve consistency with a minimal number of deletions.)

After deleting the diagonal, a consistent spin
assignment $\Sigma $ is: $\sigma _1=\sigma _3=1$ and $\sigma _2=\sigma _4=-1$, see
Figure~\ref{5-node-figure}(a).
(Another assigment is the one with all spins reversed:
$\sigma _1=\sigma _2=-1$ and $\sigma _3=\sigma _4=1$.)
If we now bring back the deleted edge, we see that in the original graph
only the one edge from node 1 to node 4 is inconsistent for the spin assignment
$\Sigma $ (Figure~\ref{5-node-figure}(c)).

This example illustrates a general fact: minimizing the number of edges that
must be removed so that there remains a consistent graph is equivalent to
finding a spin assignment $\Sigma $ for which the number of inconsistent edges
(those for which $J_{ij}\sigma _i\sigma _j=-1$) is minimized.

\begin{figure}[h,t]
 \begin{center}
 \pic{0.3}{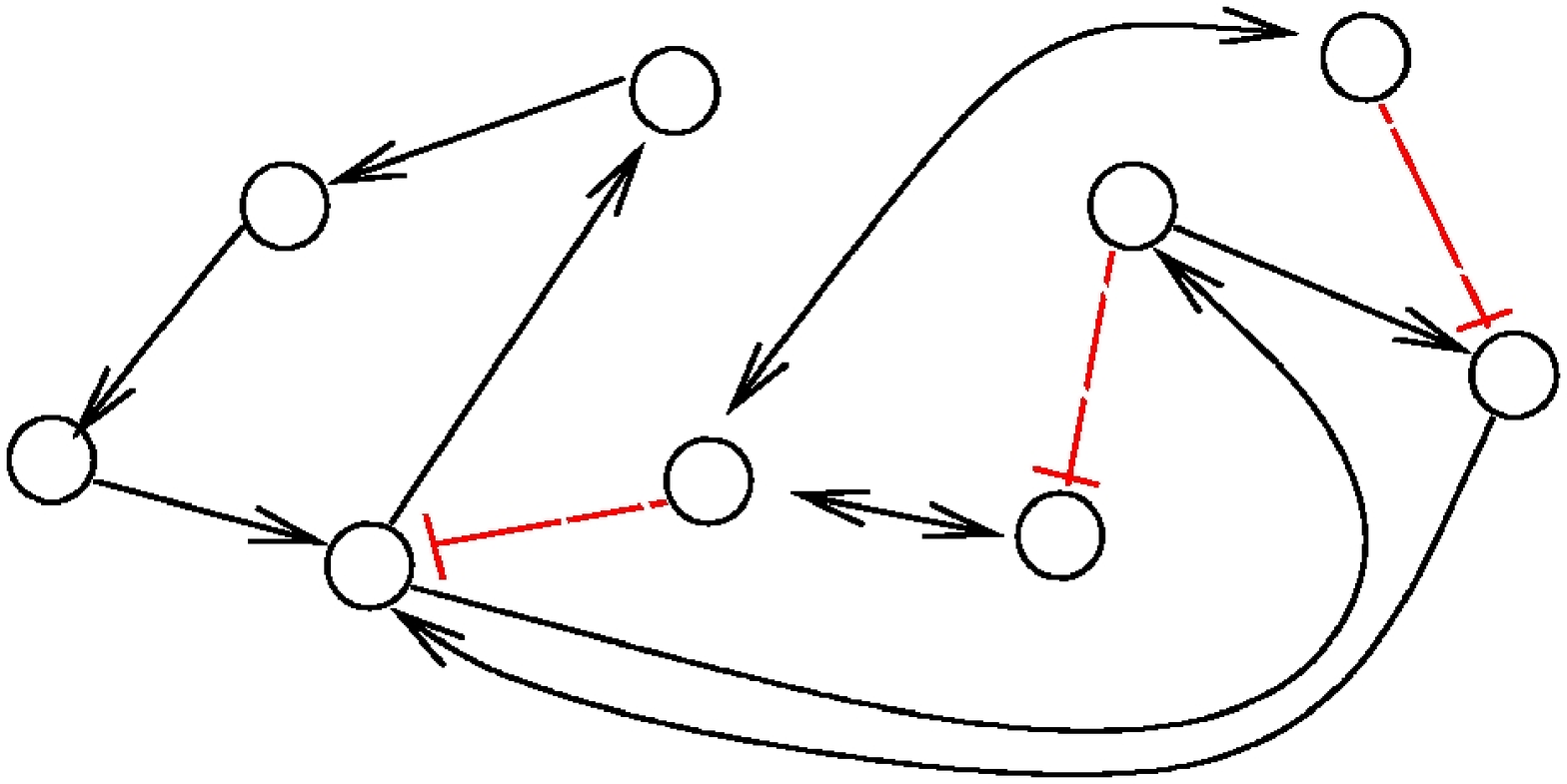}\hskip0.6cm
 \pic{0.3}{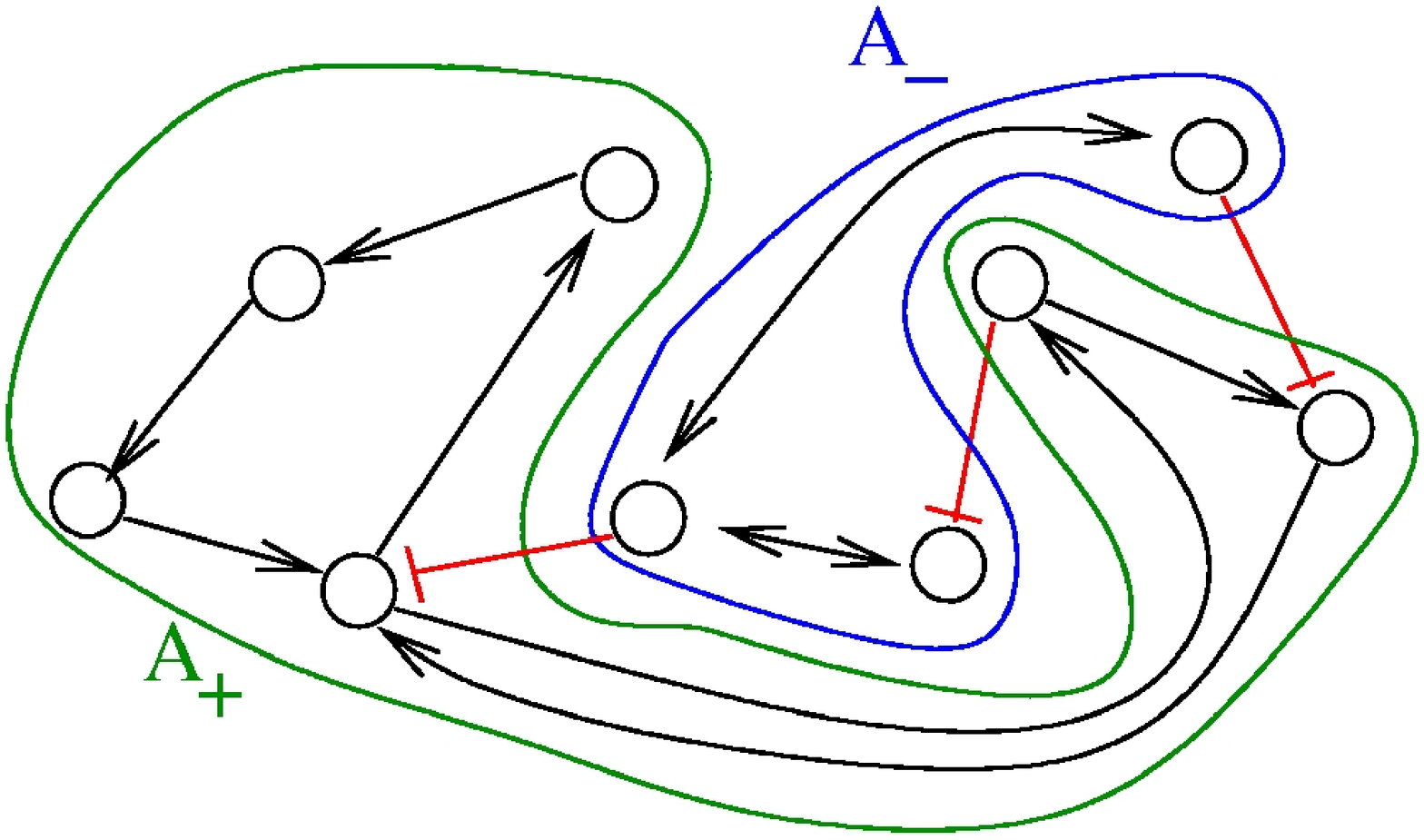}
 \caption{(a) Consistent graph; (b) partition into $A_{1}$ and $A_{-1}$}
 \label{fig:partition}
 \end{center}
\end{figure}

Yet another rephrasing is as follows.  For any spin
assignment $\Sigma $, let $A_1$ be the subset of nodes labeled $+1$, and let
$A_{-1}$ be the subset of nodes labeled $-1$.
The set of all nodes is partitioned into $A_1$ and $A_{-1}$.
(In Figure~\ref{5-node-figure}(c), we have $A_1=\{1,3\}$ and 
$A_{-1}=\{2,4\}$.)
Conversely, any partition of the set of nodes into two subsets can be thought
of as a spin assignment.
With this interpretation, a consistent spin assigment is the same as a
partition of the node set into two subsets $A_1$ and $A_{-1}$ in such a manner
that all edges between elements of $A_1$ be positive, all edges between
elements of $A_{-1}$ be positive, and all edges between a node in $A_1$ and a
node in $A_{-1}$ be negative,
see \Fig{fig:partition}.
More generally,
computing the CD amounts to finding a partition so that
$n_1+n_{-1}+p$ is minimized, where
$n_1$ is the number of negative edges between nodes in $A_1$,
$n_{-1}$ is the number of negative edges between nodes in $A_{-1}$,
and $p$ is the number of positive edges between nodes in $A_1$ and $A_{-1}$.

A very special case is when the graph has all of
its edges labeled negative, that is, $J_{ij}=-1$ for all $i,j$.
Stated in the language of partitions, the CD problem amounts to searching
for a partition such
that $n_1+n_{-1}$ is minimized (as there are no positive edges, $p=0$).
Moreover, since there are no positive edges, $n_1+n_{-1}$ is actually the
total number of edges between any two nodes in $A_1$ or in $A_{-1}$.  
Thus, $N-(n_1+n_{-1})$ is the number of remaining edges, that is, the number
of edges between nodes in $A_1$ and $A_{-1}$. 
Therefore, minimizing $n_1+n_{-1}$ is the same as maximizing $N-(n_1+n_{-1})$.
This is precisely the standard ``MAX-CUT''
problem in computer science.  

As a matter of fact, not only is MAX-CUT a
particular case, but, conversely, it is possible to reduce the CD
problem to MAX-CUT by means of the following trick.
For each edge labeled $+1$, say from $v_i$ to $v_j$, delete the edge but
insert a new node $w_{ij}$, and two negative edges,
one from $v_i$ to $w_{ij}$ and one from $w_{ij}$ to $v_j$:
\[
v_i\, \rightarrow \, v_j \quad\quad\leadsto
  \quad\quad v_i \,\dashv\, w_{ij}\, \dashv\, v_j \,.
\]
The enlarged graph has only negative edges, and it
is easy to see that the minimal number of edges that have to be removed in
order to achieve consistency is the same as the number of edges that would
have had to be removed in the original graph.  Unfortunately, the MAX-CUT
problem is NP-hard.  However, the paper~\cite{biosystems06} gave an
approximation polynomial-time algorithm for the CD problem, guaranteed to
solve the problem to within 87.9\% of the optimum value, as an adaptation
of the semi-definite programming relaxation approach to MAX-CUT based on
Goemans and Williamson's work~\cite{Goemans:1995:JACM}.  
(Is not enough to simply apply the MAX-CUT algorithm to the enlarged graph
obtained by the above trick, because the approximation bound is degraded
by the additional edges, so the construction takes some care.)

\subsubsection*{Relation to Ising spin-glass models}

Another interpretation of CD uses the language of statistical mechanics.
An \emph{Ising spin-glass model} is defined by a graph $G$
together with an ``interaction energy'' $J_{ij}$ associated to each edge
(in our conventions, $J_{ij}$ is associated to the edge from $v_j$ to
$v_i$).  In binary models, $J_{ij}\in \{1,-1\}$, as we have here.
A spin-assignment $\Sigma $ is also called a (magnetic) ``spin configuration.''
A ``non-frustrated'' spin-glass model is one for which there is a spin
configuration for which every edge is
consistent~\cite{barahona,desimone,istrail}.   
This is the same as a consistent assignment for the graph $G$ in our
terminology. 
Moreover, a spin configuration that maximizes the number of consistent edges
is one for which the ``free energy'' (with no exterior magnetic field):
\[
H(\Sigma ) \;=\; - \sum_{ij} J_{ij}\sigma _i\sigma _j
\]
is minimized.  This is because, if $\Sigma $ results in $C(\Sigma )$ consistent edges,
then $H(\Sigma ) = -C(\Sigma )+I(\Sigma ) = T-2C(\Sigma )$, where $I(\Sigma )$ is the number of
non-consistent edges for the assignment $\Sigma $ and $T=C+I$ is the total number of
edges; thus, minimizing $H(\Sigma )$ is the same as maximizing $C(\Sigma )$.
A minimizing $\Sigma $ is called a ``ground state''.
(A special case is that in which $J_{ij}=-1$ for all edges, the
``anti-ferromagnetic case''.  This is the same as the MAX-CUT problem.)

\subsubsection*{Near-monotone systems may be ``practically'' monotone}

Obviously, there is no reason for large biochemical networks to be consistent,
and they are not.
However, when the number of inconsistencies in a biological interaction graph is
small, it may well be the case that the network is in fact consistent in a
practical sense.
For example, a gene regulatory network represents all \emph{potential} effects
among genes.  These effects are often mediated by proteins which themselves
need to be activated in order to perform their function, and this activation
will, in turn, be contingent on the ``environmental'' context: extracellular
ligands, additional genes being expressed which may depend on
cell type or developmental stage, and so forth.
Thus, depending on the context, different subgraphs of the original graph
describe the system, and these graphs may be individually consistent
even if the entire graph, the union of all these subgraphs, is not.
As an illustration, take the system in Figure~\ref{consistent_fig}(c).
Suppose that under environmental conditions A, the edge from 1 to 2 is not
present, and under non-overlapping conditions B, the edge from 1 to 3 is not
be present. Then, under either conditions, A or B, the graph is
consistent, even though, formally speaking, the entire network is not
consistent. 

The closer to consistent, the more likely that this phenomenon may occur.

\subsubsection*{Some evidence suggesting near-monotonicity of natural networks}

Since consistency in biological networks may be desirable,
one might conjecture that natural biological networks tend to
be consistent.
As a way to test this hypothesis, the CD algorithm from~\cite{biosystems06}
was run on the yeast \emph{Saccharomyces cerevisiae}
gene regulatory network from~\cite{alon02}, downloaded from~\cite{yeastnet}.
(The authors of~\cite{alon02} used the YPD database~\cite{Costanzo2001}.  Nodes
represent genes, and edges are directed from transcription factors, or
protein complexes of transcription factors, into the genes regulated by them.)
This network has 690 nodes and 1082 edges, of which 221 are negative
and 861 are positive
(we labeled the one ``neutral'' edge as positive; the conclusions do not
change substantially if we label it negative instead, or if we delete this one
edge).
The algorithm in~\cite{biosystems06} 
provides a CD of 43.
In other words, \emph{deleting a mere 4\% of edges makes the network
consistent}.
Also remarkable is the following fact.  The original graph has 11 components:
a large one of size 664, one of size 5, three of size 3, and six of size 2.
All of these components \emph{remain connected} after edge deletion.  The
deleted edges are all from the largest component, and they are incident on a
total of 65 nodes in this component.

To better appreciate if a small CD might happen by chance, the algorithm was
also run on random graphs having 690 nodes and 1082 edges (chosen uniformly),
of which 221 edges (chosen uniformly) are negative.
It was found that, for such random graphs, about 12.6\% 
($136.6\pm 5$) of edges have to be removed in order to achieve consistency.  
(To analyze the scaling of this estimate, we generated random graphs with $N$
nodes and $1.57N$ edges of which $0.32N$ are negative.  We found that for
$N>10$, approximately $N/5$ nodes must be removed, thus confirming the result
for $N=690$.) 
Thus, the CD of the biological network is roughly 15
standard deviations away from the mean for random graphs.
Both topology (i.e., the underlying graph) and
actual signs of edges contribute to this near-consistency of the yeast
network.  To justify this assertion, the following numerical experiment
was performed.  We randomly changed the signs of 50 positive and 50 negative
edges, thus obtaining a network that has the same number of positive and
negative edges, and the same underlying graph, as the original yeast network,
but with 100 edges, picked randomly, having different signs.  Now, one needs
8.2\% ($88.3\pm7.1$) deletions, an amount in-between that obtained for the
original yeast network and the one obtained for random graphs.  Changing more
signs, 100 positives and 100 negatives, leads to a less consistent network,
with $115.4\pm4.0$ required deletions, or about 10.7\% of the original edges,
although still not as many as for a random network.

\subsubsection*{Decomposing systems into monotone components}

\begin{figure}[h,t]
\begin{center}
\setlength{\unitlength}{1800sp}%
\begin{picture}(3024,1449)(3289,-2998)
\thicklines
\put(3601,-2161){\framebox(2400,600){}}
\put(6001,-1861){\line( 1, 0){300}}
\put(6301,-1861){\line( 0,-1){900}}
\put(6301,-2761){\vector(-1, 0){900}}
\put(3301,-1861){\vector( 1, 0){300}}
\put(3301,-1861){\line( 0,-1){900}}
\put(3301,-2761){\line( 1, 0){900}}
\put(4201,-2986){\framebox(1200,450){}}
\put(4000,-2000){consistent}%
\put(4500,-2900){``$-$''}%
\end{picture}
\caption{Pulling-out inconsistent connections}
\label{pull-out}
\end{center}
\end{figure}
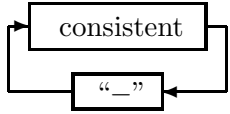

Another motivation for the study of near-monotone systems is from
decomposition-based methods for the analysis of systems that are
interconnections of monotone subsystems.
One may ``pull out'' inconsistent connections among monotone components, in
such a manner that the original system can then be viewed as a ``negative
feedback'' loop around an otherwise consistent system (Figure~\ref{pull-out}).
In this interpretation, the number of interconnections among monotone
components corresponds to the number of variables being fed-back.

For example, let us take the graph shown in Figure~\ref{5-node-figure}(a).
The procedure of dropping the diagonal edge and seeing it instead as an
external feedback loop can be modeled as follows.
The original differential equation $\fractextnp{dx_1}{dt}=f_1(x_1,x_2,x_3,x_4)$
is replaced by the equation $\fractextnp{dx_1}{dt}=f_1(x_1,x_2,x_3,u)$, where
the symbol $u$, which represents an external input signal, is inserted instead
of the state variable $x_4$.
The consistent system in
Figure~\ref{pull-out} includes the remaining four edges, and the ``negative''
feedback (negative in the sense that it is inconsistent with the rest of the
system) is the connection from $x_4$, seen as an ``output'' variable, back
into the input channel represented by $u$.  The closed-loop system obtained by
using this feedback is the original system, now viewed as a negative feedback
around the consistent system in Figure~\ref{5-node-figure}(b). 

Generally speaking, the decomposition techniques
in~\cite{monotoneTAC,monotoneMulti,pnasangeliferrellsontag04,sysbio04,ejc05es,angelileenheersontagSCL04,predatorpreysgt05,enciso_sontagSCL05,leenheer-malisoff,dcds06,enciso_smith_sontagJDE06,gedeon05}
are most useful if the feedback loop involves few variables.
This is equivalent to asking that
the graph $G$ associated to the system be close to
consistent, in the sense of the CD of $G$ being small.
This view of systems as monotone systems ---which have strong stability
properties, as discussed next,---  with negative-feedback regulatory loops
around them is very appealing from a control engineering perspective as well.

\subsubsection{Dynamical behavior of monotone systems}

Continuous-time monotone systems have convergent behavior.  For example,
they cannot admit any possible stable oscillations
\cite{hadeler83,hirschAMS84,Hirsch-Smith}.
When there is only one steady state, a theorem of Dancer \cite{dancer98}
shows --under mild assumptions regarding possible constraints on the values of
the variables, and boundedness of solutions-- that every solution converges to
this unique steady state (monostability).
When, instead, there are multiple steady-states, the Hirsch Generic
Convergence Theorem \cite{Hirsch,Hirsch2,smith,Hirsch-Smith} is the
fundamental result.  A \emph{strongly} monotone system is one for which the 
an initial perturbation $z_i(0)>x_i(0)$ on the concentration of any species
propagates as a strict up or down perturbation: 
$z_j(t)>x_j(t)$ for all $t>0$ and all indices $j$ for which $\sigma _j=\sigma _i$,
and $z_j(t)<x_j(t)$ for all $t>0$ and all $j$ for which $\sigma _j=-\sigma _i$.
Observe that this requirement is stronger (hence the terminology) than merely
weak inequalities: $z_j(t)\geq x_j(t)$ or $z_j(t)\leq x_j(t)$ respectively as in
Kamke's Theorem.
A sufficient condition for strong monotonicity is that the Jacobian matrices
must be irreducible for all $x$, which basically amounts to asking that the
graph $G$ must be strongly connected and that every non-identically zero
Jacobian entry be everywhere nonzero.
Even though they may have arbitrarily large dimensionality, monotone
systems behave in many ways like one-dimensional systems:
Hirsch's Theorem asserts that generic bounded solutions of strongly monotone
differential equation systems must converge to the set of steady states.
(``Generic'' means ``every solution except for a measure-zero set of initial
conditions.'')  In particular, no ``chaotic'' or other ``strange''
dynamics can occur.
For discrete-time strongly monotone systems, generically also stable
oscillations are allowed besides convergence to equilibria, but no more
complicated behavior. 

The ordered behavior of monotone systems is robust with respect to spatial
localization effects as well as signaling delays (such as those arising from
transport, transcription, or translation).  Moreover, their stability
character does not change much if some inconsistent connections are inserted,
but only provided that these added connections are weak 
(``small gain theorem'') or
that they operate at a comparatively fast time scale
\cite{06posta_wang}.

\begin{figure}[h,t]
 \begin{center}
\pic{0.4}{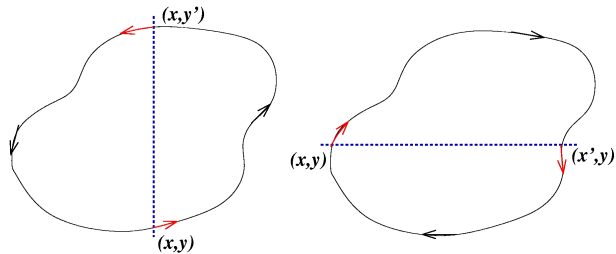}
 \caption{Impossible (a) counterclockwise and (b) clockwise
periodic orbits in planar cooperative system, each drawn in the $(x,y)$-plane.}
 \label{contradict_cooperative}
 \end{center}
\end{figure}

The intuition behind the convergence results is easy to explain in
the very special case of just two interacting species, described by
a two-dimensional system with variables $x(t)$ and $y(t)$:
\beqn
\frac{dx}{dt}&=&f(x,y)\\
\frac{dy}{dt}&=&g(x,y) \,.
\eeqn
A system like this is monotone if either (a) the species are mutually
activating (or, as is said in mathematical biology, ``cooperative''), (b)
they are mutually inhibiting (``competitive''), or (c) either $x$ does not
affect $y$, $y$ does not affect $x$, or neither affects the other.
Let us discuss the mutually activating case (a).  
(Case (b) is similar, and
case (c) is easy, since the systems are partially or totally decoupled.)
We want to argue that there cannot be any periodic orbit.
Suppose that there would be a periodic orbit in which the motion is
counterclockwise, as shown in Figure \ref{contradict_cooperative}(a).
We then pick two points in this orbit with identical $x$ coordinates, as
indicated by $(x,y)$ and $(x,y')$ in Figure \ref{contradict_cooperative}(a).
These points correspond to the concentrations at two times $t_0$, $t_1$,
with $x(t_0)=x(t_1)$ and $y(t_0)<y(t_1)$.
Since $y(t_1)$ is larger than $y(t_0)$, $x$ is at the same concentration, and
the species are mutually activating, it follows that the rate of change in the
concentration $x$ should be comparatively larger at time $t_1$ than at time
$t_0$, that is, $f(x,y')\geq f(x,y)$.
However, this contradicts the fact that $x(t)$ is increasing at time $t_0$ 
($f(x,y)\geq 0$) but is decreasing at time $t_1$ ($f(x,y')\leq 0$).
The contradiction means that there cannot be any counterclockwise-oriented
curve.  To show that there cannot be any clockwise-oriented
curve, one may proceed by an entirely analogous argument, using two
points $(x,y)$ and $(x',y)$ as in Figure \ref{contradict_cooperative}(b).
Of course, the power of monotone systems theory arises in the analysis of
systems of higher dimension, since two-dimensional systems are
easy to study by elementary phase plane methods.

For general, non-monotone systems, on the other hand, no dynamical
behavior, including chaos, can be mathematically ruled out.
This is in spite of the fact that some features of non-monotone systems are
commonly regarded as having a stabilizing effect.  For example, negative
feedback loops confer robustness with regard to certain types of structural as
well as external perturbations 
\cite{doyle90,
sepulchre97,
montreal98,
khalil02}%
.
However, and perhaps paradoxically, the behavior of non-monotone systems may
also be very fragile: for instance, they can be destabilized by delays in
negative feedback paths.
Nonetheless, we conjecture that systems that are close to monotone must be
better-behaved, generically, than those that are far from monotone.
Preliminary evidence (unpublished) for this has been obtained from the
analysis of random Boolean networks, at least for discrete analogs of the
continuous system, but the work is not yet definitive.

\subsubsection*{Directed cycles}

Intuition suggests that somewhat less than monotonicity should suffice for
guaranteeing that no chaotic behavior may arise, or even that no stable limit
cycles exist.  Indeed, monotonicity amounts to requiring that no undirected
negative-parity cycles be present in the graph, but a weaker condition, that
no \emph{directed} negative parity
cycles exist, should be sufficient to insure these
properties.  For a strongly connected graph, the property that no 
directed negative cycles exist is equivalent to the property that no undirected
negative cycles exist, because the same proof as given earlier, but applied to
directed paths, insures that a consistent spin assignment exists (and hence
there cannot be any undirected negative cycles).
However, for non-strongly connected graphs, the properties are not the same.
On the other hand, every graph can be decomposed as a cascade of graphs that
are strongly connected.
This means (aside from some technicalities having to do with Jacobian entries
being not identically zero but vanishing on large sets) that systems having no 
directed negative cycles can be written as a cascade of strongly monotone
systems.  Therefore, it is natural to conjecture that such cascades have
nice dynamical properties.
Indeed, under appropriate technical conditions for the systems in the cascade,
one may recursively prove convergence to equilibria in each component,
appealing to the theory of asymptotically autonomous systems \cite{thieme92}
and thus one may conclude global convergence of the entire system
\cite{hirsch-neural89,smith-neural91}.
For example, a cascade of the form $dx/dt=f(x)$, $dy/dt=g(x,y)$ where the $x$
system is monotone and
where
the system $dy/dt=g(x_0,y)$ is monotone for each fixed
$x_0$, cannot have any attractive periodic orbits (except equilibria).
This is because the projection of such an orbit on the first system must be
a point $x_0$, and hence the orbit must have the form $(x_0,y(t))$.
Therefore, it is an attractive periodic orbit of $dy/dt=g(x_0,y)$, and by
monotonicity of this latter system we conclude that $y(t)\equiv $ a constant as
well. 
The argument generalizes to any cascade, by an inductive argument.
Also, chaotic attractors cannot exist \cite{angeli_hirsch_sontag}.

The condition of having no directed negative cycles is the weakest one that
can be 
given strictly on the basis of the graph $G$, because for any graph $G$ with
a negative feedback loop there is a system with graph $G$ which admits stable
periodic orbits.  (First find a limit cycle for the loop, and then use a small
perturbation to define a system with nonzero entries as needed, which will
still have a limit cycle.)

\subsubsection*{Positive feedback and stability}

The strong global convergence properties of monotone systems mentioned above
would seemingly contradict the fact that positive feedback, which tends to
increase the 
direction of perturbations, is allowed in monotone systems, but negative
feedback, which tends to stabilize systems, is not.
One explanation for this apparent paradox is that the main theorems in
monotone systems theory only guarantee that \emph{bounded} solutions converge,
but they do not make any assertions about unbounded solutions.  For example,
the system $dx/dt = -x + x^2$ has the
property that every solution starting at an $x(0)>1$ is unbounded, diverging to
$+\infty $, a fact which does not contradict its monotonicity (every
one-dimensional system is monotone).
This is not as important a restriction as it may seem, because for biochemical
systems it is often the case that all trajectories must remain bounded, due to
conservation of mass and other constraints.
A second explanation is that negative \emph{self-loops} are not ruled out in
monotone systems, and such loops, which represent degradation or decay
diagonal terms, help insure stability.

\subsubsection*{Intuition on why negative self-loops do not affect monotonicity}

In the definition of the graph associated to a continuous-time system,
self-loops (diagonal terms in the Jacobian of the vector field $f$) were
ignored.  The theory (Kamke's condition) does not require self-loop
information in order to guarantee monotonicity.  Intuitively, the reason for
this is that a larger initial value for a variable $x_i$ implies a larger
value for this variable, at least for short enough time periods, independently
of the sign of the partial derivative $df_i/dx_i$ (continuity of flow with
respect to initial conditions).  For example, consider a degradation equation
$dp/dt = -p$, for the concentration $p(t)$ of a protein P.  At any time $t$,
we have that $p(t)=e^{-t}p(0)$, where $p(0)$ is the initial concentration.
The concentration $p(t)$ is positively proportional to $p(0)$, even though the
partial derivative $\frac{\partial (-p)}{\partial p}=-1$ is negative.  Note
that, in contrast, for a difference equation, a jump may occur: for instance
the iteration $p(t+1)=-p(t)$ has the property that the order of two elements
is reversed at each time step.  Thus, for difference equations, diagonal terms
matter.

\subsubsection*{Multiple time scale analysis may make systems monotone}

A system may fail to be monotone due to the effect of negative regulatory
loops that operate at a faster time scale than monotone subsystems.  In such a
case, sometimes an approximate but monotone model may be obtained, by
collapsing negative loops into self-loops.
Mathematically: a non-monotone system might be a singular perturbation
of a monotone system.
A trivial linear example that illustrates this point
is
$dx/dt$$\,=\,$$-x$$-y$, $\varepsilon dy/dt$$\,=\,$$-y$$+$$x$, with $\varepsilon $$>$$0$.
This system is not monotone (with respect to any orthant cone).
On the other hand, for $\varepsilon \ll1$, the fast variable $y$ tracks $x$,
so the slow dynamics is well-approximated by $dx/dt=-2x$
(monotone, since every scalar system is).

More generally, one may consider $dx/dt=f(x,y)$, $\varepsilon dy/dt=g(x,y)$
such that the fast system $dy/dt=g(x,y)$ 
has a unique globally asymptotically stable steady state $y=h(x)$ for each
$x$ (and possibly a mild input to state stability requirement, as with
the special case $\varepsilon dy/dt=-y+h(x)$),
and the slow system $dx/dt=f(x,h(x))$ is (strongly) monotone.
Then one may expect that the original system inherits
global convergence properties, at least for all $\varepsilon $$>$$0$ small enough.
The paper~\cite{06cdc_wang_sontag} employs tools from
geometric
invariant manifold theory
\cite{Fenichel,Jones}, taking advantage of the existence of
a manifold $M_\varepsilon $ invariant for the dynamics, which attracts all near-enough
solutions, and with an asymptotic phase property.
The system restricted to the invariant manifold $M_\varepsilon $ is a regular
perturbation of the fast ($\varepsilon =0$) system, and hence inherits strong
monotonicity properties.  So, solutions in the manifold will be generally
well-behaved, and asymptotic phase implies that solutions track
solutions in $M_\varepsilon $, and hence also converge to equilibria if
solutions on $M_\varepsilon $ do.
However, the technical details are delicate, because strong monotonicity
only guarantees generic convergence, and one must show
that the generic tracking solutions start from the ``good'' set of initial
conditions, for generic solutions of the large system.

\subsubsection*{Discrete-time systems}

As discussed, for autonomous differential equations monotonicity implies
that stable periodic behaviors will not be observed, and moreover, under
certain technical assumptions, all trajectories must converge to steady states.
This is not exactly true for difference equation models, but a variant 
does hold: for discrete-time monotone systems,
trajectories must converge to either steady states \emph{or periodic orbits}.
In general, even the simplest difference equations may exhibit
arbitrarily complicated (chaotic) behavior, as shown by the logistic iteration
in one dimension $x(t+1)=kx(t)(1-x(t))$ for appropriate values of
the parameter $k$ \cite{devaney89}. 
However, for monotone difference equations, a close analog of Hirsch's Generic
Convergence Theorem is known.
Specifically, suppose that the equations are point-dissipative, meaning that
all solutions converge to a bounded set \cite{hale88},
and that the system is strongly monotone, in the sense that the Jacobian
matrix $(\partial f_i / \partial x_j)$ is irreducible at all states.
Then, a result of Tere\v s\v c\'ak and coworkers
\cite{PolTer92,PolTer93,hess93,terec96}
shows that there is a positive integer $m$ such that generic solutions
(in an appropriate sense of genericity) converge to periodic orbits with
period at most $m$.  
Results also exist under less than strong monotonicity, just as in the
continuous case, for example when steady states are unique \cite{dancer98}.

Difference equations allow one to study wider classes of
systems.  As a simple example, consider the nondimensionalized harmonic
oscillator (idealized mass-spring system with no damping), which has
equations
\beqn
\frac{dx}{dt}&=&y\\
\frac{dy}{dt}&=&-x \,.
\eeqn
(For this example, we allow variables to be negative; these variables might
indicate deviations of concentrations from some reference value.)
This system is not monotone, since $v_1\rightarrow v_2$ is negative and $v_1\rightarrow v_2$ is
positive, so that its graph has a negative loop.
On the other hand, suppose that one looks at this system every $\Delta t$ seconds,
where $\Delta t=\pi $.  The discrete-time system that results (using a superscript
${}^+$ to indicate time-stepping) is now:
\beqn
x^+&=&-y\\
y^+&=&-x
\eeqn
(this is obtained by solving the differential equation on an interval of
length $\pi $).
This system is monotone (both $v_1\rightarrow v_2$ and $v_1\rightarrow v_2$ are negative).
Every trajectory of this discrete system is, in fact, of period two:
$(x_0,y_0)\rightarrow (-y_0,-x_0)\rightarrow (x_0,y_0)\rightarrow \ldots $.
This periodic property for the difference equation corresponds to the
period-2$\pi $ behavior of the original differential equation.

\subsubsection*{Oscillatory behaviors}

Stable periodic behaviors are ruled-out in autonomous monotone continuous-time
systems.  However, stable periodic orbits may arise through various external 
mechanisms.
Three examples are (1) inhibitory negative feedback from some species into
others in a monotone monostable system, (2) the generation of relaxation
oscillations from a hysteresis parametric behavior by negative feedback on 
parameters by species in a monotone system, and (3) entrainment of external
periodic signals.
These general mechanisms are classical and well-understood for simple, one or
two-dimensional, dynamics, and they may be generalized to the case where the
underlying system is higher-dimensional but monotone.

\subsubsection*{Embeddings in monotone systems}

\begin{figure}[h,t]
 \begin{center}
 \pic{0.25}{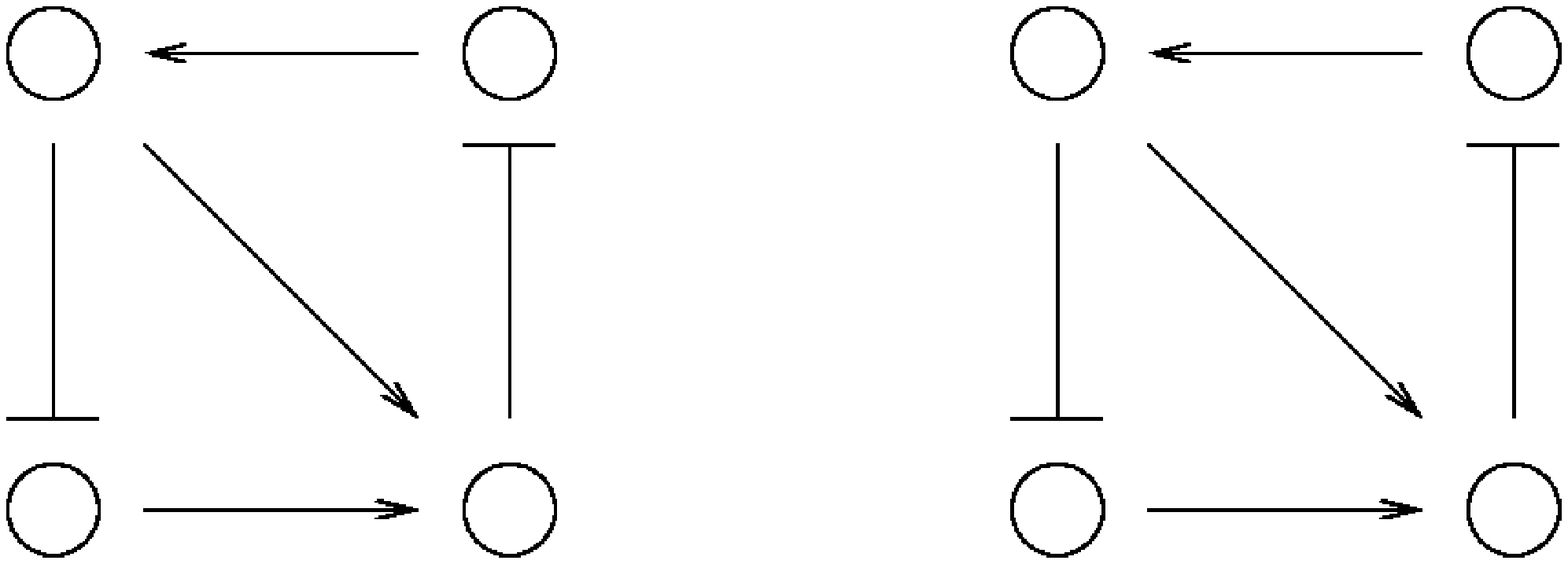}
\hskip2cm
 \pic{0.25}{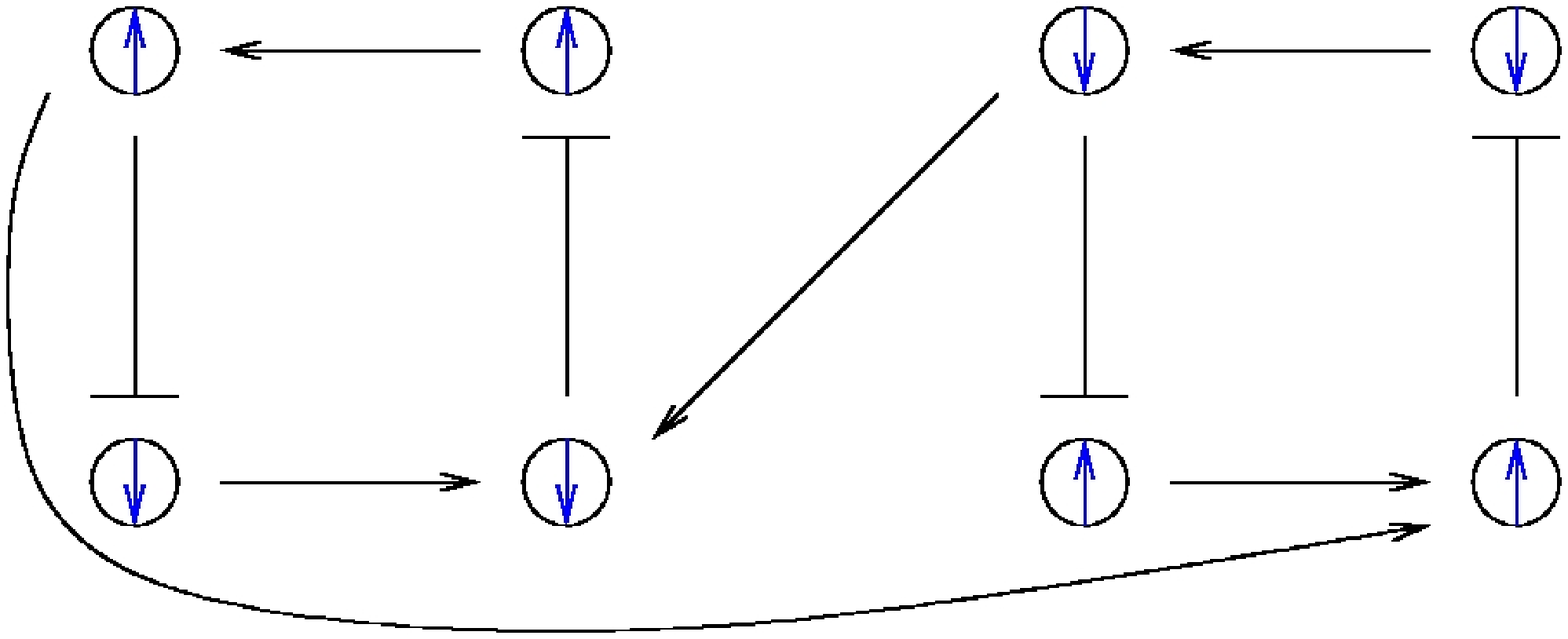}
 \caption{(a) duplicated inconsistent graph, (b) replacing arrows and
consistent assignment}
 \label{duplicating_graph}
 \end{center}
\end{figure}

As observed by Gouz\'e \cite{gouzeINRIA88,gouze_hadeler94},
any $n$-dimensional system
can be viewed as a subsystem of a $2n$-dimensional monotone system.
The mathematical trick is to first duplicate every variable
(species), introducing a ``dual'' species, and then to replace every
inconsistent edge by an edge connecting the source species and the ``dual'' of
its target 
(and vice-versa).
The construction is illustrated in \Fig{duplicating_graph}.
At first, this embedding result may seem paradoxical, since all
monotone (or strongly monotone) systems have especially nice dynamical
behaviors, such as not having any attractive periodic orbits or chaotic
attractors, and of course non-monotone systems may admit such behaviors. 
However, there is no contradiction.  A non-monotone subsystem of a monotone
system may well have, say, a chaotic attractor or a stable periodic orbit: it
is just that this attractor or orbit will be unstable when seen as a subset
of the extended ($2n$-dimensional) state space.  
Not only there is no contradiction, but a classical construction of
Smale~\cite{smale76} shows that indeed any possible dynamics can be embedded
in a larger monotone system.
More generally, the Hirsch Generic Covergence Theorem guarantees convergence
to equilibria from \emph{almost every} initial condition; applied to the above
construction, in general the exceptional set of initial conditions would
include the ``thin'' set corresponding to the embedded subsystem.
Yet, one may ask what happens for example if the larger $2n$-system has a
unique equilibrium.  In that case, it is known~\cite{dancer98} that every
trajectory converges (not merely generic ones), so, in particular, the
embedded subsystem must also be ``well-behaved''.  Thus, systems that may be
embedded by the above trick into monotone systems with unique equilibria will
have global convergence to equilibria.  This property amounts to the ``small
gain theorem'' shown in~\cite{monotoneTAC},
see~\cite{enciso_smith_sontagJDE06} for a discussion and further results using
this embedding idea. 

\subsubsection*{Discrete systems}

We remark that one may also study difference equations for which the state
components 
are only allowed to take values out of a finite set.  For example,
in Boolean models of biological networks, each variable $x_i(t)$
can only attain two values ($0/1$ or ``on/off'').  These values represent
whether the $i$th gene is being expressed, or the concentration of the $i$th
protein is above certain threshold, at time $t$.
When detailed information on kinetic rates of protein-DNA or protein-protein
interactions is lacking, and especially if regulatory relationships are
strongly sigmoidal, such models are useful in theoretical analysis,
because they serve to focus attention on the basic dynamical characteristics
while ignoring specifics of reaction mechanisms
\cite{kauffman69a,%
kauffman69b,kauffman-glass73,albertothmer03,chaves_albert_sontagJTB05}).

For difference equations over finite sets, such as Boolean systems,
it is quite clear that all trajectories must either settle into equilibria or
to periodic orbits, whether the system is monotone or not.
However, cycles in discrete systems may be arbitrarily long and these might be
seen  as ``chaotic''  motions.   Monotone systems,  while  also settling  into
steady states or periodic orbits, have generally shorter cycles.
This is because periodic orbits must be anti-chains, i.e.\ no two different
states can be compared; see~\cite{Gilbert54,smith}.
For example, consider a discrete-time system in which species concentrations
are quantized to the $k$ values $\{0,\ldots ,\kmo\}$;
we interpret monotonicity with respect to the partial order:
$(a_1,\ldots ,a_n)\leq (b_1,\ldots ,b_n)$ if every coordinate $a_i\leq b_i$.
For non-monotone systems, orbits can have as many as $k^n$ states.
On the other hand, monotone systems cannot have orbits of size more than
the width (size of largest antichains)
of $P=\{0,\ldots ,\kmo\}^n$, which can be interpreted as the set of
multisubsets of an $n$-element set, or equivalently as the set of divisors of 
a number of the form $(p_1p_2\ldots p_n)^{\kmo}$ where the $p_i$'s are distinct
primes.  The width of $P$ is the number of possible vectors $(i_1,\ldots ,i_n)$
such that $\sum i_j = \lfloor{kn/2}\rfloor$ and each $i_j\in \{0,\ldots ,\kmo\}$.
This is a generalization of Sperner's Theorem; see \cite{andersonsperner}.
For example, for $n=2$, periodic orbits in a monotone system evolving on
$\{0,\ldots ,\kmo\}^2$ cannot have length larger than $k$, while non-monotone
systems on $\{0,\ldots ,\kmo\}^2$ can have a periodic orbit of period $k^2$.
As another example, arbitrary Boolean systems (i.e., the state space is
$\{0,1\}^n$) can have orbits of period up to $2^n$, but monotone systems
cannot have orbits of size larger than 
${n \choose \lfloor{n/2}\rfloor}\approx 2^n \sqrt{2 / (n \pi)}$.
These are all classical facts in Boolean circuit design~\cite{Gilbert54}.
It is worth pointing out that any anti-chain $P_0$ can be seen as a periodic
orbit of a monotone system.
This is proved as follows: we enumerate the elements of $P_0$ as
$x_1,\ldots ,x_\ell$, and define $f(x_i)=x_{i-1}$ for all $i$ modulo $\ell$.
Then, $f$ can be extended to all elements of the state space by defining
$f(x)=(0,\ldots ,0)$ for every $x$ which has the property that $x<x_i$
for some $x_i\in P_0$ and $f(x)=(\kmo,\ldots ,\kmo)$ for every $x$ which is not
$\leq x_i$ for any $x_i\in P_0$.
It is easy to see that this is a monotone map \cite{Gilbert54,aracena04}.

While on the subject of discrete and in particular Boolean systems,
we mention a puzzling fact: any Boolean function may be implemented by using
just two inverters, with all other gates being monotone.
In other words, a circuit computing any Boolean rule whatsoever may be
built so that its ``consistency deficit'' is just two.
This is a well-known fact in circuit design~\cite{Gilbert54,Minsky67}.
Here is one solution, from \cite{GeorgeHarper}.
One first shows how to implement the Boolean function
that takes as inputs three bits $A,B,C$ and outputs the vector of three 
complements $(\mbox{not} A, \mbox{not} B, \mbox{not} C)$, by using this
sequence of operations: 
\beqn
2or3ones &=& (A \wedge B) \vee (A \wedge C) \vee (B \wedge C)\\
0or1ones &=& \mbox{not} (2or3ones)\\
1one     &=& 0or1ones \wedge (A \vee B \vee C)\\
1or3ones &=& 1one \vee (A \wedge B \wedge C)\\
0or2ones &=& \mbox{not} (1or3ones)\\
0ones    &=& 0or2ones \wedge 0or1ones\\
2ones    &=& 0or2ones \wedge 2or3ones\\
\mbox{not}A &=& 0ones \vee (1one \wedge (B \vee C)) 
  \vee (2ones \wedge (B \wedge C))\\
\mbox{not}B &=& 0ones \vee (1one \wedge (A \vee C)) 
  \vee (2ones \wedge (A \wedge C))\\
\mbox{not}C &=& 0ones \vee (1one \wedge (A \vee B))
  \vee (2ones \wedge (A \wedge B))
\eeqn
(the node labeled ``2or3ones'' computes the Boolean function ``the input has
exactly 2 or 3 ones'' and so forth).
Note that only two inverters have been used.
If we now want to invert four bits $A,B,C,D$, we build the above circuit, but
we implement the inversion of the three bits
$(2or3ones,1or3ones,D)$ by a subciruit with only two inverters.
With a similar recursive construction, one may invert an arbitrary number of
bits, using just two inverters.

\bigskip
(continued in Part 2)

\end{document}